\documentclass{aa}

\usepackage{graphicx,textcomp}
\usepackage[varg]{txfonts}
\usepackage[colorlinks=true, linkcolor=blue, citecolor=blue]{hyperref}
\usepackage{multirow}
\usepackage{threeparttable}
\usepackage{natbib}
\usepackage{pdflscape}
\usepackage{rotating}
\usepackage{tabularx}
\usepackage{xcolor}

\usepackage[version=4]{mhchem} 
\bibpunct{(}{)}{;}{a}{}{,} 
\usepackage{twoopt}
\hypersetup{
  colorlinks,
  citecolor=[rgb]{0.06, 0.2, 0.65},
  linkcolor=[rgb]{0.8, 0.2, 1.0},
  urlcolor=[rgb]{0.06, 0.2, 0.65},
}
\setcitestyle{maxnames=2} 

\input{macros.tex}

\begin{document}

   \title{Searching for Molecular Signatures in 14 Transiting Exoplanets with SPIRou}

   \author{A. Masson\inst{1}, 
           S. Vinatier\inst{2},
           B. Bézard\inst{2},
           F. Debras\inst{3},
           A. Carmona\inst{3},
           J. Lillo-Box\inst{1},
           N. B. Cowan\inst{4},
           V. Yariv\inst{5},
           R. Allart\inst{6}
          }
   \institute{1. Centro de Astrobiología (CAB), CSIC-INTA, Camino Bajo del Castillo s/n, 28692, Villanueva de la Cañada, Madrid, Spain\\
              2. LIRA, Observatoire de Paris, Université PSL, Sorbonne Université, Université Paris Cité, CY Cergy Paris Université, CNRS, 92195 Meudon, France\\
              3. Institut de Recherche en Astrophysique et Planétologie, Université de Toulouse, CNRS UMR 5277, 14 avenue Edouard Belin, 31400 Toulouse, France\\
              4. Department of Physics and Department of Earth \& Planetary Sciences, McGill University, 3550 rue University, Montréal, QC H3A 2A7, Canada\\
              5. Institut de Planétologie et Astrophysique de Grenoble, Grenoble, CNRS, IPAG, 38000 Grenoble, France\\
              6. D\'epartement de Physique, Institut Trottier de Recherche sur les Exoplan\`etes, Universit\'e de Montr\'eal, Montr\'eal, Qu\'ebec, H3T 1J4, Canada\\ \\
              \email{\href{mailto:amasson@cab.inta-csic.es}{amasson@cab.inta-csic.es} - \href{mailto:amasson.astro@gmail.com}{amasson.astro@gmail.com}}\\
                      }
   \date{Received Month dd, yyyy; accepted Month dd, yyyy}

  \titlerunning{Searching for Molecular Signatures in 14 Transiting Exoplanets with SPIRou}
  \authorrunning{A. Masson et al.}

  \abstract{High-resolution spectroscopy provides unique constraints on exoplanet atmospheric composition and dynamics. The past decade of ground-based campaigns has accumulated extensive public archives, yet many observations remain unanalyzed. We present a homogeneous blind-search analysis of 50 SPIRou transits spanning 14 exoplanets, ranging from super-Earths to ultra-hot Jupiters, combining data from large program and public archived observations. Using automated data reduction and atmospheric retrieval via Nested Sampling validated by Cross Correlation Function analysis, we confirm previous \ce{H2O} and \ce{CO} detections in HD\,189733\,b, WASP-76\,b, and WASP-127\,b, report tentative \ce{H2S} detections in HD\,189733\,b and TOI-1807\,b, tentative detection of GJ\,3470\,b's atmosphere, and provide upper limits for non-detections. This work demonstrates a scalable method for systematic archive analysis, providing a first step toward ground-based support of large space-based atmospheric characterization programs and the study of atmospheric diversity across exoplanet populations from a statistical perspective.}


  \maketitle

  \section{Introduction}
  Ground-based high-resolution spectroscopy (HRS, with a typical resolving power $\mathcal{R}\gtrsim25\,000$) has emerged as a powerful method for exoplanetary atmospheric characterization. Transit observations provide access to atmospheric composition and thermal structure in planetary limbs through transmission spectroscopy \citep[e.g.,][]{Snellen2008,Snellen2010,Birkby2013,Brogi2018,Boucher2021}. These constraints in turn provide valuable information on planetary formation pathways and migration processes in the protoplanetary disk, particularly from key elemental ratios such as O/H, C/H, and C/O \citep[e.g.,][]{Oberg2011,Madhusudhan2011,Moses2013a,Marboeuf2014,Helling2014}. 
  
  The high resolving power of ground‑based spectrographs, compared with lower‑resolution space‑based observations, further enables Doppler‑shift measurements of atmospheric dynamics and probing of \red{lower-pressure} levels less affected by cloud and haze opacity, making HRS a key complement to space‑based observations \citep[e.g.,][]{Gandhi2020}. As JWST rapidly expands the sample of characterized atmospheres and Ariel is expected to eventually add hundreds more \citep{Tinetti2018,Turrini2022}, ground-based HRS support is needed to provide complementary wavelength and pressure levels coverage and enable population-level studies through homogeneous data analysis methodology.
  
  Atmospheric characterization constitutes a primary science objective for SPIRou, a near-infrared spectropolarimeter mounted on the 3.6\,m Canada-France-Hawaii Telescope (CFHT). Since its first light in 2018, three coordinated international large programs---the SPIRou Legacy Survey (SLS), SPICE (SPIRou Legacy Survey -- Consolidation and Enhancement), and ATMOSPHERIX---have enabled successful \ce{H2O} and \ce{CO} abundances measurements in HD\,189733\,b \citep{Boucher2021,Klein2024} and WASP-76\,b \citep{Hood2024}, alongside atmospheric escape detections via the metastable helium triplet \citep{Allart2023,Masson2024}. However, many targets observed within these large programs received limited per-object transit coverage due to allocation constraints, which has so far prevented robust atmospheric retrievals and left many datasets unexploited. With complementary observations now publicly available, comprehensive cross-program reanalysis becomes feasible. We present a homogeneous blind-search atmospheric analysis of 14 SPIRou transiting exoplanets spanning super-Earths to ultra-hot Jupiters, using 50 individual transits from SLS, SPICE, ATMOSPHERIX, and public archived data acquired by SPIRou. 
  Section\,\ref{observations} presents the observations. Section\,\ref{reduction} describes our data reduction and processing methodology, while Sec.\,\ref{method} details the atmospheric modeling and retrieval approach. Results for each target are presented and discussed in Sec.\,\ref{result}, with Sec.\,\ref{conclusion} summarizing our conclusions and future prospects for HRS atmospheric characterization.

  \section{Observations}
  \label{observations}
  SPIRou's instrumental capabilities are detailed in \citet{Donati2018,Donati2020}; we here summarize the specifications relevant to this work. Each exposure covers a continuous spectral range between 0.95 and 2.50\mum with a resolving power of 70\,000 and a sampling precision of $\sim$2.25\kms per spectral bin. The instrument operates in a cryogenic vessel at 80\,K stabilized to $<$2\,mK precision. Data are distributed over 49 spectral orders, indexed from \#31 in the red to \#79 in the blue, each sampled over 4088 spectral bins. As in \citet{Masson2024}, we focused our analysis on the 15 transiting short-period exoplanets observed within the SLS, SPICE, and ATMOSPHERIX large programs. Our target list hence contains 2 super-Earths (55\,Cnc\,e, GJ\,486\,b), 2 sub-Neptunes (GJ\,1214\,b, TOI-1807\,b), 5 warm Neptunes (AU\,Mic\,b, GJ\,436\,b, GJ\,3470\,b, HAT-P-11\,b, K2-25\,b), 2 hot Saturn (WASP-69\,b, WASP-127\,b), 1 warm Jupiter (WASP-80\,b), 2 hot Jupiters (HD\,189733\,b, HD\,209458\,b), and 1 ultra-hot Jupiter (WASP-76\,b). K2-25\,b's single transit exhibited extremely low SNR \citep{Allart2023,Masson2024} and was discarded, reducing our sample to 14 targets. We compiled a total of 50 transits: 10 from SLS/SPICE (PI: J.-F. Donati), 16 from ATMOSPHERIX \citep[PI: F. Debras;][]{Debras2024,Klein2024}, and 24 from public archives. The complete list of observations is detailed in Table\,\ref{transit_tab}. Data are available via the CADC database\footnote{\url{https://www.cadc-ccda.hia-iha.nrc-cnrc.gc.ca/}} once publicly released. Throughout this analysis, we employed the same stellar and planetary parameters as in \citet{Masson2024} and refer to their Tables\,\href{https://www.aanda.org/articles/aa/full_html/2024/08/aa49608-24/T2.html}{2} and \href{https://www.aanda.org/articles/aa/full_html/2024/08/aa49608-24/T3.html}{3} and references therein.
  
  \begin{table*}[]
  \centering
	\begin{threeparttable}
 	\caption[Transit characteristics of the studied targets]{Transit characteristics of the studied targets}
 	\label{transit_tab}
 	\begin{tabular}{lllllllll}
  	\hline \hline
  	Target                    & Program ID  & PI                    & UT date     	& Midpoint               	& Duration            	& Texp        & Nexp       & DRS version  \\
                              &             &                     	& [yyyy-mm-dd]	& [\BJD]              	  & [hours]               & [sec]       &            &              \\ 
  	\hline
    55 CnC e                 	& 19AC11    	& E. Deibert          	& 2019-02-14  	&          2459213.92319	& 1.59                 	&         95	&        77  & 0.7.291      \\ \vspace{1pt}
    55 CnC e                 	& 19AC11    	& E. Deibert          	& 2019-02-25  	&          2459213.92319	& 1.59                 	&         95	&        73  & 0.7.291      \\ \vspace{1pt}
    55 CnC e                 	& 19AC11    	& E. Deibert          	& 2019-04-17  	&          2459213.92319	& 1.59                 	&         95	&        54  & 0.7.291      \\ \vspace{1pt}
    55 CnC e                 	& 19AC11    	& E. Deibert          	& 2019-05-01  	&          2459213.92319	& 1.59                 	&         95	&        73  & 0.7.291      \\ \vspace{1pt}    
    55 CnC e                 	& 20BF09    	& F. Debras           	& 2020-12-30  	&          2459213.92319	& 1.59                 	&         45	&        176 & 0.7.291      \\ \vspace{1pt}
    AU Mic b                 	& 19AP42    	& J.-F. Donati        	& 2019-06-17  	&          2458651.98451	& 1.66                 	&        123	&        116 & 0.7.291      \\ \vspace{1pt}    
    GJ 1214 b                	& 19AP40    	& J.-F. Donati        	& 2019-04-18  	&          2458983.91348	& 0.86                 	&        602	&        11  & 0.7.294      \\ \vspace{1pt}
    GJ 1214 b                	& 20AP40    	& J.-F. Donati        	& 2020-05-14  	&          2458983.91348	& 0.86                 	&        251	&        30  & 0.7.294      \\ \vspace{1pt}
    GJ 1214 b                	& 23BF16    	& F. Debras           	& 2023-09-21  	&          2458983.91348	& 0.86                 	&        407	&        15  & 0.7.294      \\ \vspace{1pt}
    GJ 1214 b                	& 24AF11    	& F. Debras           	& 2024-03-22  	&          2458983.91348	& 0.86                 	&        407	&        16  & 0.7.294      \\ \vspace{1pt}
    GJ 1214 b                	& 24AF11    	& F. Debras           	& 2024-04-21  	&          2458983.91348	& 0.86                 	&        407	&        16  & 0.7.294      \\ \vspace{1pt}    
    GJ 3470 b                	& 19AP40    	& J.-F. Donati        	& 2019-02-18  	&          2458532.90622	& 1.39                 	&        435	&        32  & 0.7.294      \\ \vspace{1pt}
    GJ 3470 b                	& 21BP40    	& J.-F. Donati        	& 2021-12-15  	&          2458532.90622	& 1.39                 	&        301	&        43  & 0.7.294      \\ \vspace{1pt}
    GJ 3470 b                	& 22BF19    	& F. Debras           	& 2023-01-06  	&          2458532.90622	& 1.39                 	&        501	&        7   & 0.7.294      \\ \vspace{1pt}
    GJ 3470 b                	& 23BF16    	& F. Debras           	& 2023-12-29  	&          2458532.90622	& 1.39                 	&        552	&        20  & 0.7.294      \\ \vspace{1pt}
    GJ 3470 b                	& 24AF11    	& F. Debras           	& 2024-04-04  	&          2458532.90622	& 1.39                 	&        552	&        20  & 0.7.294      \\ \vspace{1pt}    
    GJ 436 b                 	& 19AP40    	& J.-F. Donati        	& 2019-02-25  	&          2458540.10182	& 0.72                 	&        217	&        22  & 0.7.294      \\ \vspace{1pt}    
    GJ 486 b                 	& 21AD97    	& F. Debras           	& 2021-03-24  	&          2459297.93910	& 0.97                 	&        217	&        30  & 0.7.294      \\ \vspace{1pt}
    GJ 486 b                 	& 21BC34    	& E. Deibert          	& 2022-01-23  	&          2459297.93910	& 0.97                 	&        184	&        42  & 0.7.294      \\ \vspace{1pt}
    GJ 486 b                 	& 22AS06    	& W. Wang             	& 2022-03-11  	&          2459297.93910	& 0.97                 	&        212	&        47  & 0.7.288      \\ \vspace{1pt}
    GJ 486 b                 	& 22AS06    	& W. Wang             	& 2022-03-14  	&          2459297.93910	& 0.97                 	&        212	&        46  & 0.7.288      \\ \vspace{1pt}
    GJ 486 b                 	& 22AS06    	& W. Wang             	& 2022-04-11  	&          2459297.93910	& 0.97                 	&        212	&        46  & 0.7.288      \\ \vspace{1pt}    
    HAT-P-11 b               	& 21AF18    	& F. Debras           	& 2021-06-30  	&          2459395.93748	& 2.41                 	&        323	&        44  & 0.7.294      \\ \vspace{1pt}
    HAT-P-11 b               	& 21BF19    	& F. Debras           	& 2021-08-18  	&          2459395.93748	& 2.41                 	&        323	&        44  & 0.7.294      \\ \vspace{1pt}
    HAT-P-11 b               	& 21BC09    	& M. Radica           	& 2021-08-13  	&          2459395.93748	& 2.41                 	&        284	&        46  & 0.7.288      \\ \vspace{1pt}
    HAT-P-11 b               	& 22AF11    	& F. Debras           	& 2022-06-12  	&          2459395.93748	& 2.41                 	&        323	&        37  & 0.7.294      \\ \vspace{1pt}    
    HD 189733 b              	& 18BD50    	& C. team             	& 2018-09-22  	&          2458383.80116	& 1.97                 	&        251	&        36  & 0.7.294      \\ \vspace{1pt}
    HD 189733 b              	& 19AP40    	& J.-F. Donati        	& 2019-06-15  	&          2458383.80116	& 1.97                 	&        251	&        50  & 0.7.294      \\ \vspace{1pt}
    HD 189733 b              	& 20AC01    	& E. Deibert          	& 2020-07-03  	&          2458383.80116	& 1.97                 	&         95	&        54  & 0.7.294      \\ \vspace{1pt} 
    HD 189733 b              	& 20AC01    	& E. Deibert          	& 2020-07-05  	&          2458383.80116	& 1.97                 	&         95	&        52  & 0.7.294      \\ \vspace{1pt} 
    HD 189733 b              	& 20AC01    	& E. Deibert          	& 2020-07-25  	&          2458383.80116	& 1.97                 	&         95	&        124 & 0.7.294      \\ \vspace{1pt} 
    HD 189733 b              	& 21BC16    	& E. Deibert          	& 2021-08-24  	&          2458383.80116	& 1.97                 	&         95	&        108 & 0.7.291      \\ \vspace{1pt} 
    HD 189733 b              	& 23BO37    	& A. Oklopcic         	& 2023-08-04  	&          2458383.80116	& 1.97                 	&         95	&        84  & 0.7.291      \\ \vspace{1pt} 
    HD 189733 b              	& 23BO37    	& A. Oklopcic         	& 2023-08-24  	&          2458383.80116	& 1.97                 	&         95	&        80  & 0.7.291      \\ \vspace{1pt} 
    HD 189733 b              	& 24AF11    	& F. Debras           	& 2024-05-25  	&          2458383.80116	& 1.97                 	&        301	&        30  & 0.7.294      \\ \vspace{1pt}
    HD 189733 b              	& 24AC16    	& R. Allart           	& 2024-06-25  	&          2458383.80116	& 1.97                 	&         95	&        112 & 0.7.290      \\ \vspace{1pt}     
    HD 209458 b              	& 19AF16    	& G. Hebrard          	& 2019-06-18  	&          2458653.03998	& 3.09                 	&        903	&        18  & 0.7.294      \\ \vspace{1pt}                    
    TOI-1807 b              	& 21BD87    	& B. Klein            	& 2022-01-12  	&          2459684.94696	& 1.00                 	&        362	&        20  & 0.7.291      \\ \vspace{1pt}
    TOI-1807 b               	& 22AF11    	& F. Debras           	& 2022-04-15  	&          2459684.94696	& 1.00                 	&        362	&        20  & 0.7.294      \\ \vspace{1pt}
    TOI-1807 b               	& 22AF11    	& F. Debras           	& 2022-06-09  	&          2459684.94696	& 1.00                 	&        362	&        20  & 0.7.294      \\ \vspace{1pt}    
    WASP-127 b               	& 20AP42    	& J.-F. Donati        	& 2020-03-11  	&          2458919.96815	& 4.54                 	&        301	&        50  & 0.7.294      \\ \vspace{1pt}
    WASP-127 b               	& 21AC02    	& A. Boucher          	& 2021-03-22  	&          2458919.96815	& 4.54                 	&        501	&        28  & 0.7.294      \\ \vspace{1pt}
    WASP-127 b               	& 21AC02    	& A. Boucher          	& 2021-05-03  	&          2458919.96815	& 4.54                 	&        501	&        20  & 0.7.294      \\ \vspace{1pt}
    WASP-127 b              	& 23BC26    	& A. Boucher          	& 2023-12-28  	&          2458919.96815	& 4.54                 	&        501	&        30  & 0.7.288      \\ \vspace{1pt}
    WASP-127 b              	& 23BC26    	& A. Boucher          	& 2024-01-22  	&          2458919.96815	& 4.54                 	&        501	&        30  & 0.7.288      \\ \vspace{1pt}    
    WASP-69 b                	& 19BP40    	& J.-F. Donati        	& 2019-10-13  	&          2458769.85111	& 2.23                 	&        123	&        93  & 0.7.294      \\ \vspace{1pt}
    WASP-76 b                	& 20BF09    	& F. Debras           	& 2020-10-31  	&          2459153.88490	& 3.75                 	&        774	&        27  & 0.7.288      \\ \vspace{1pt}
    WASP-76 b                	& 21BF19    	& F. Debras           	& 2021-10-28  	&          2459153.88490	& 3.75                 	&        774	&        26  & 0.7.294      \\ \vspace{1pt}
    WASP-76 b                	& 21BS19    	& G. Chen             	& 2021-09-18  	&          2459153.88490	& 3.75                 	&        301	&        61  & 0.7.288      \\ \vspace{1pt}
    WASP-80 b                	& 19BP40    	& J.-F. Donati        	& 2019-10-07  	&          2458763.77156	& 1.27                 	&        184	&        74  & 0.7.291      \\
  	\hline
 	\end{tabular}

 	\begin{tablenotes}[flushleft]
    	\footnotesize{\item{\textbf{Notes.} Transit midpoints were computed using the periods and midpoint reference values from the ExoClock\tablefootmark{*} database \citep{Kokori2022_a,Kokori2022_b,Kokori2023}. All dates are defined in Barycentric Julian Date (BJD) expressed in the Dynamic Barycentric Time standard (TDB), the proper time definition for defining interstellar events \citep{Eastman2010}. "Duration" is the transit duration, in hours. "Texp" is the exposure time, in seconds, of a single exposure. "Nexp" is the total number of exposures collected during the transit. "DRS version" is the version of the Data Reduction Software used to pre-process the data (see Sec.\,\ref{reduction}). \\
    	\tablefootmark{*}{\url{https://www.exoclock.space/}}
    	}}
 	\end{tablenotes}

   \end{threeparttable}
  \end{table*}
  
  \section{Data Reduction}
  \label{reduction} 

  \subsection{Data preparation with \texttt{APERO}}
  Ground-based high-resolution spectroscopy of exoplanet atmospheres requires careful calibration and removal of telluric and stellar contaminants to extract the faint planetary signal. SPIRou data reduction for atmospheric characterization follows established methodologies documented in recent studies \citep{Deibert2021,Boucher2021,Klein2024,Debras2024,Masson2024,Hood2024}. We present a brief overview of our independent reduction pipeline\footnote{Available at \url{https://github.com/admasson/HR-SpARTA}}, which applies standard HRS techniques validated through injection-recovery tests.
  We start with 2D (spectral order, wavelength) echelle data produced by \texttt{APERO} \citep{Artigau2018}, the SPIRou Data Reduction Software (DRS). These files are wavelength-calibrated (defined in vacuum through this study) and corrected for detector effects (bad pixels, instrumental background, nonlinearities, cosmic rays) and telluric transmission \citep{Cook2022}. Signal-to-noise ratios (\snr) are measured at 1.65 µm from photon counts. Since most observations were obtained in non-polarimetric mode, we used the combined flux throughout. For each transit, we used data produced with the latest DRS version available\footnote{We used data processed at the SPIRou Data Centre hosted at the Laboratoire d’Astrophysique de Marseille for DRS versions 0.7.294 or higher, complemented, when unavailable, with data from the CADC processed with DRS version 0.7.288.} for that specific observation date (Table~\ref{transit_tab}). Although this required combining transits produced with different DRS versions for some targets, we found that improvements in telluric correction from newer DRS versions justified this approach.

  \subsection{Reduction workflow}
  \label{reduction_workflow}
  Using the \texttt{APERO} produced data, our reduction workflow first correct for instrumental blaze using profiles stored in the data file. Most telluric absorption is already handled by the DRS, but some residuals remain, and the bins corresponding to the strongest corrected Earth lines exhibit large correlated noise. We therefore apply telluric lines clipping following \citet{Boucher2021} by flagging telluric lines with core transmission below 0.4 and masking spectral bins from line core up to 0.97 transmission in the wings. Each order spectrum is normalized by its median along the spectral axis, followed by an iterative 3-$\sigma$ clipping using 2nd-order polynomial time-series models for each wavelength bin with a maximum of five iterations. We flatten the continuum by dividing each order by its moving average computed with a 119-pixel kernel\footnote{The \texttt{convolve} method from \texttt{astropy.convolution}, which we used, requires an odd kernel.} to remove low-frequency instrumental modal noise. Data are then shifted to the stellar rest frame using $V = V_{\mathrm{sys}} - \mathrm{BERV}$, where $V_{\mathrm{sys}}$ is the systemic velocity and BERV the Barycentric Earth Radial Velocity \citep{Wright2014}. We remove the stellar contribution by dividing each exposure by a master spectrum built, for each transit, from an (\snr)$^2$-weighted average of both in-transit and out-of-transit observations. Including in-transit data was required to robustly estimate and correct the mean stellar spectrum. Although these data may contain planetary signal, we estimate that, for each target, the planet’s radial velocity (RV) shift during transit is large enough for this contribution to average out in the master spectrum. Our tests showed that performing this step in the stellar rest-frame resulted in a better correction of the stellar spectral signatures and extraction of the planetary signal despite the additional noise introduced when shifting and spectraly interpolating the data. Following \citet{Masson2024}, a (\snr)$^2$ weight was applied to any temporal combinations of the data to optimally account for the \snr temporal variations.

  \subsection{Additional correction with SysRem}
  Stellar template division removes only the temporal mean stellar signature, leaving time-variable components uncorrected. Residual variations arise from changing observing conditions, stellar activity, \red{stellar line distortions induced by center-to-limb variations (CLV; see \citealp{Dravins2017}) or the} Rossiter-McLaughlin effect (RME; \citealp{McLaughlin1924}), and imperfect telluric correction. We remove these percent-level residuals using the SysRem algorithm \citep{Tamuz2005,Mazeh2007}, which was reported \citep{Birkby2013,Birkby2017,Ridden2023,Nortmann2025} to perform such correction equivalently or better than Principal Component Analysis (PCA). SysRem iteratively identifies basis vectors maximizing projected data variance while accounting for uncertainties, then subtracts projections along these principal vectors to remove correlated noise. Performing K iterations of SysRem is therefore conceptually similar to removing the K first Principal Components (PC) in a PCA-based approach, with the added advantage that SysRem accounts for varying uncertainties in the data. 

  \begin{figure}[!htbp]
    \centering
    \includegraphics[width=\hsize,trim=1cm 0cm 1cm 1cm,clip]{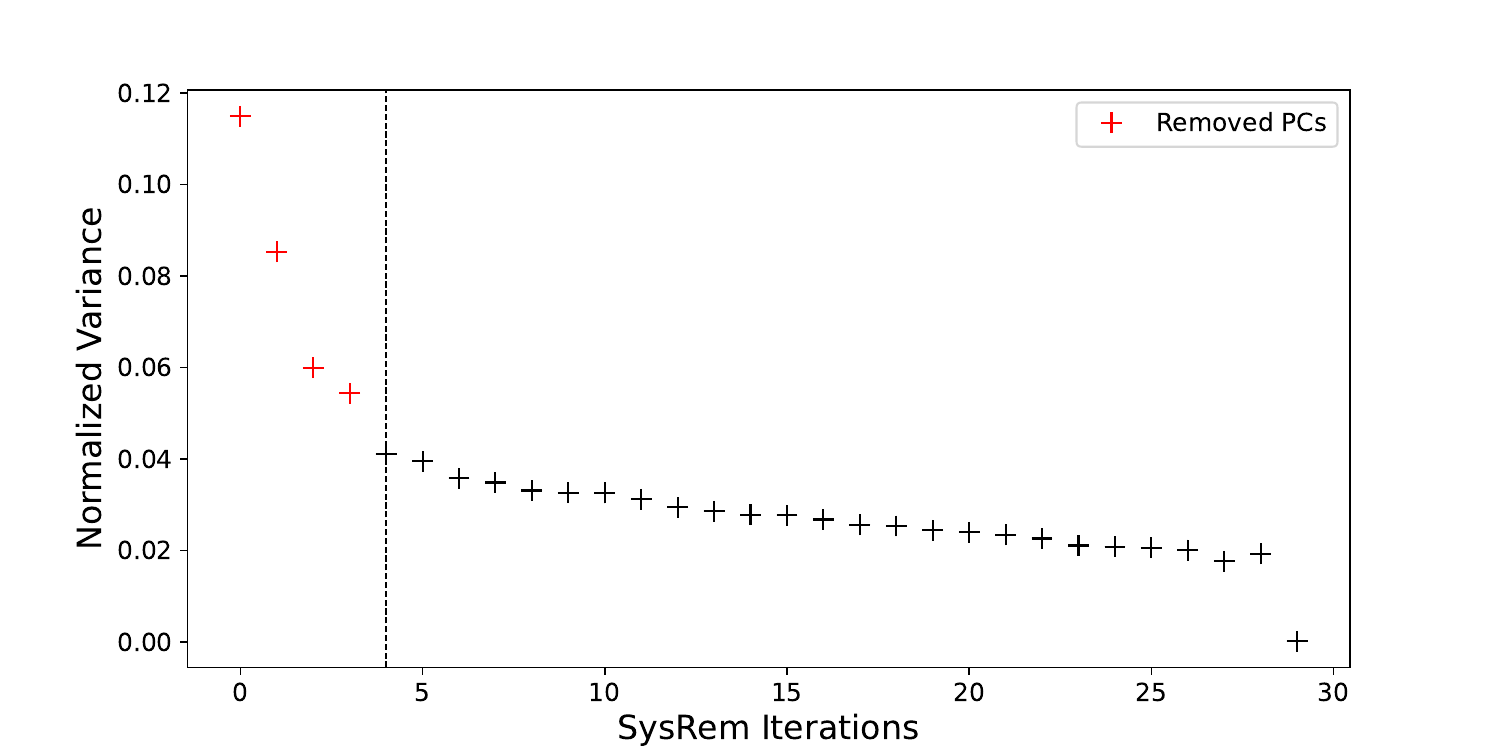}
    \caption{Example of the data variance explained per SysRem iteration for order \#68 of the WASP-127\,b 2023 transit. The vertical dashed line marks the Kneedle-identified elbow giving the optimal number of SysRem iterations/PCs to remove (red crosses). The final zero-variance point$-$due to the dimensionality reduction caused by the mean stellar template removal$-$was excluded from the Kneedle analysis.}
    \label{fig:scree_plot}
  \end{figure}

  Determining the optimal number of iterations per spectral order is critical: too few iterations leave correlated residuals, while too many may suppress the planetary signal. A common approach is to tune this number by maximizing the recovered planet signal, but this method has been shown to potentially bias and artificially increase the resulting \snr \citep{Cabot2019,Cheverall2023}. The data variance explained per SysRem iteration follows a linear trend for pure Gaussian noise but exhibits an "elbow" shape in real data, where early iterations remove high-variance systematics before transitioning to a noise-dominated regime (Fig.\ref{fig:scree_plot}). We employ the Kneedle\footnote{\url{https://github.com/arvkevi/kneed}} algorithm \citep{Satopaa2011}$-$designed to detect elbow positions in 1D curves$-$to automatically determine the optimal number of SysRem iteration per order. This approach is faster, less biased than methods maximizing output planetary signals, and allows for blind automation of residual noise removal in the observed data. We validated our approach with injection-recovery tests and found that i) SysRem slightly outperforms PCA, ii) applying the method to the data in log-flux space slightly improves performance, and iii) optimizing the per-order number of SysRem iterations with Kneedle yielded better correction. Following this noise removal step with SysRem, a final sigma-clipping stage removes any remaining outliers based on spectral variance trends (Appendix\,\ref{std_sigma_clipping}). All reduced data obtained from the above steps are publicly accessible on the associated online repository\footnote{\url{https://cloud.cab.inta-csic.es/s/a3G7DzExPtaF7L7}}.

  \section{Method}
  \label{method}
  \subsection{Modeling the transmission signature}
  \label{NS_method}
  We compute atmospheric transmission spectra using \texttt{petitRADTRANS} \citep{Molliere2019,Molliere2020,Alei2022,Blain2024,Nasedkin2024}, which performs 1D radiative transfer given planetary bulk parameters, thermal profiles, and composition. The code accounts for H$_2$ and He Rayleigh scattering, H$_2$-H$_2$ and H$_2$-He collision-induced absorptions, and optically thick gray cloud decks at a given pressure level. We assumed hydrogen-dominated\footnote{Although 55\,Cnc\,e, GJ\,486\,b, and TOI-1807\,b may not host primary atmospheres (in which case the retrieved parameters would be physically non-informative), we retain an \ce{H2}-dominated assumption for all targets to keep the blind-search analysis homogeneous.} atmospheres and \red{adopted a free-chemistry approach by fitting constant-with-height mass mixing ratios (MMRs) for a dozen of species (see Section\,\ref{NS_section}). For a given composition, petitRADTRANS then automatically fills the remaining fraction with \ce{H2} and He at a fixed 75/25 mass ratio and computes the corresponding mean molecular weight self-consistently.} We used a pressure grid of 130 layers logarithmically spaced from $10^{-8}$ to $10^{2}$ bar, and used the line-by-line mode to compute the wavelength-dependent apparent planetary radius $R_\lambda$ at a resolving power of $10^6$. We reduced computation time by setting petitRADTRANS' wavelength sampling parameter to 4, reducing the model's sampling while maintaining at least 2 times oversampling relative to the data. 
  
  We then construct time-varying transmission spectra \mbox{$T_\lambda(t) = 1 - W(t) (R_\lambda/R_\star)^2$}, where $R_\star$ is the host star radius, and $W(t)$ is the transit light curve computed via \texttt{batman} \citep{Kreidberg2015} taking into account orbital parameters and quadratic limb darkening. Rotational broadening due to atmospheric rotation is applied \red{following the convolution-based methodology of \citet{Klein2024}}, then spectra are shifted to the stellar rest frame using the \texttt{KeplerEllipse} model from \texttt{pyastronomy} to account for orbital geometry and eccentricity. The model is then sampled onto SPIRou wavelength bins and orders, convolved at the instrumental resolution, and broadened for each exposure to account for the smearing induced by the planet's radial-velocity shift during exposures. We must then apply the same degradation to the planetary transmission model as induced by our reduction pipeline. For the moving average division, we divide each synthetic spectral order by its smoothed version using an identical kernel size (Section~\ref{reduction_workflow}). For the systematic noise removal with SysRem, we follow \citet{Gibson2022}: during reduction, we precomputed and stored the $\mathbf{U.U^\dagger}$ product for each spectral order, where $\mathbf{U}$ contains the first left singular vectors from each SysRem iteration and $\mathbf{U^\dagger}$ is its Moore-Penrose inverse. We thus multiply each spectral order by the corresponding $\mathbf{U.U^\dagger}$ product to apply SysRem's signal removal effect to the model with minimal computational overhead.

  \subsection{Nested Sampling}
  \label{NS_section}
  We employed Nested Sampling via \texttt{pymultinest} \citep{Buchner2014,Feroz2008,Feroz2019} for atmospheric parameter inference, chosen for its computational efficiency relative to MCMC \citep{Klein2024}. \red{We used the following likelihood expression from \citet{Gibson2020}, which allow for optimal combination of spectra when their absolute noise scale is unknown and their relative uncertainties are estimated from the data statistics:}
  \begin{equation}
    \centering
    \label{logL}
      \red{ \ln \mathcal{L} = -\frac{N}{2}\ln\!\left(\frac{1}{N}\sum_i \frac{(d_i-m_i)^2}{\sigma_i^2}\right) },
  \end{equation}

  \noindent where $d_i$ is the reduced data, $m_i$ the model, $\sigma_i$ the uncertainty, and $i$ sums over wavelength and time. As in \citet{Masson2024}, the uncertainty were estimated by computing the (\snr)²-weighted standard deviation of each spectral bin along the time axis. 
  For multi-transit targets, we co-added log-likelihoods across all transits. We restricted retrievals to isothermal atmospheres with uniform abundances due to computational constraints, searching for \ce{H2O}, \ce{CO}, \ce{CH4}, \ce{NH3}, \ce{CO2}, \ce{H2S}, \ce{HCN}, \ce{C2H2}, and \ce{OH} (opacity sources in Table~\ref{tab:opacity_sources}). We adopted a two-stage retrieval approach to handle the distinct requirements of detections versus non-detections. The \textit{Full Run} optimized 15 free parameters (Kp, $V_0$, $T_{\text{iso}}$, abundances in MMRs, \red{equatorial rotational velocity} $V_{\text{rot}}$, and cloud pressure level $P_{\text{cl}}$) to obtain constraints on clearly identified planetary signatures. However, when no clear detection exists, high-dimensional parameter degeneracies and residual noise cause the algorithm to converge toward spurious stellar or telluric features in the (Kp, $V_0$) plane rather than finding realistic atmospheric constraints. We therefore performed an \textit{Upper Limits Run} with all parameters held fixed except for molecular abundances, to prevent spurious convergence. For this run, the temperature profile was set to non-isothermal using the \texttt{cloud\textunderscore500} theoretical grid from Exo-REM\footnote{\url{https://gitlab.obspm.fr/Exoplanet-Atmospheres-LESIA/exorem}}, a 1D radiative-convective equilibrium model that simulates planetary atmospheres by accounting for cloud formation and chemical disequilibrium effects \citep{Baudino2015,Baudino2017,Charnay2018,Blain2021}. For this run, we also fixed Kp to its planetary values, $V_0$ to $0\,$\kms, 
  and excluded rotational broadening and clouds. Both runs used 400 live points. Constraints are then derived from the posterior distributions: for converged runs we report the median and the 16th–84th (1$\sigma$) percentile interval, while for non-converged runs we report the 99.7th percentile as the 3-$\sigma$ upper limit.

  \subsection{Cross Correlation Function}
  Nested Sampling can be sensitive to convergence on telluric or stellar residuals, so we employed the Cross Correlation Function (CCF) technique as an independent verification method. For each potential detection, and using same notation as Eq.\,(\ref{logL}), we computed $ CCF = \sum_{i}\frac{d_i m_i}{\sigma_i^2}$ \citep{Snellen2010}, evaluated as a function of Kp and V$_0$ by shifting the planetary model in the stellar rest frame. We converted CCF maps to \snr maps by dividing each value by the standard deviation of the CCF in a region far from the expected planetary signature, following standard practice \citep[e.g.,][]{Boucher2021,Klein2024,Hood2024}. To properly estimate the CCF variance, we sampled 500 Kp values (0--500~\kms) and 200 V$_0$ values (-100$-$100~\kms), excluding a masked region of $\pm100$~\kms in Kp and $\pm25$~\kms in V$_0$ around the expected planetary position. Our detection criterion requires: (i) Nested Sampling convergence to a solution consistent with the planetary Kp and V$_0$, and (ii) a CCF signature with SNR $\geq 5$ at that location. Signatures with SNR in the [3--5] range were flagged as tentative.

  \section{Results}
  \label{result}
  

  \begin{figure*}[!htbp]
    \centering
    \includegraphics[width=0.95\textwidth]{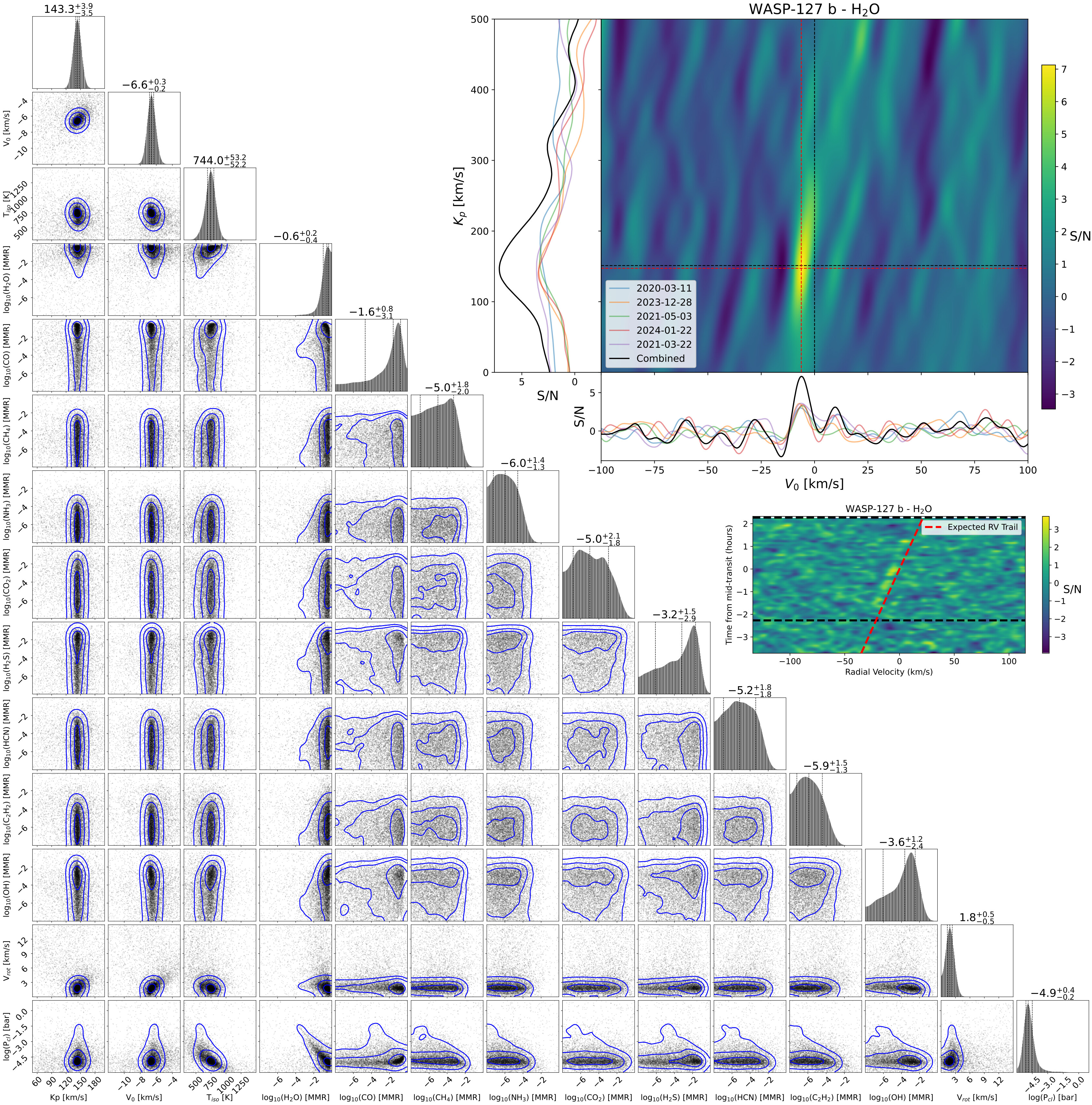}
    
    \caption{Results for WASP-127\,b. \textbf{Bottom Left:} Corner plot from the \textit{Full Run} showing 1-, 2-, and 3-$\sigma$ confidence contours (blue) with median posterior values and 1-$\sigma$ errors above each distribution. \textbf{Upper Right:} \red{CCF map of \ce{H2O} in WASP-127\,b, obtained with a model generated using the Section~\ref{NS_method} setup, with parameters set to their median posterior values.} The expected planetary position is marked with a black cross; the best-fit retrieval position (Kp, V$_0$) is shown as a red cross. Left and lower panels display the projected CCF at the best (Kp, V$_0$) for each transit (color-coded) and their combination (black). A phase‑resolved CCF is also shown, displaying the planetary \ce{H2O} signature detected during transit (horizontal dashed black line) and following the expected planetary RV trail in the stellar rest frame (red dashed line).}
    \label{fig:wasp127_ccf}
  \end{figure*}
 
  \subsection{55 Cnc e}
  \label{55cnce}
  Discovered in 2004 \citep{McArthur2004,Dawson2010}, this $\sim1.8$\,\Rearth, $\sim8$\,\Mearth super-Earth \citep{Bourrier2018a} may host a hot, possibly lava-driven secondary atmosphere \citep{Ridden-Harper2016,Demory2016,Tabernero2021,Deibert2021,Mercier2022,Keles2022,Rasmussen2023,Hu2024}. HCN was tentatively reported with HST/WFC3 \citep{Tsiaras2016}, while recent JWST observations allow for a \ce{CO2}-rich atmosphere \citep{Hu2024}. We reanalysed the four SPIRou transits from \citet{Deibert2021} plus one ATMOSPHERIX transit (Table~\ref{transit_tab}). The \textit{Full Run} converged on residuals without constraining parameters, and the \textit{Upper Limits Run} yielded unrealistically high \ce{CO2} abundances at the prior's edge (Fig.~\ref{fig:55cnce_aumic_ccf}), making this parameter uninformative. Computing the CCF map corresponding to the best retrieved abundances results in a marginal \red{3.0}-$\sigma$ peak at the expected planetary (Kp, V$_0$) position, driven primarily by \ce{CO2}. A similar marginal signal was noted by \citet{Deibert2021} without conclusive interpretation. Given the modest sensitivity of five SPIRou transits for detecting 55\,Cnc\,e's expected confined atmosphere, this signal is most likely spurious noise or telluric contamination. We hence report no atmospheric detection, consistent with prior non-detections of a primary H$_2$/He envelope \citep{Ehrenreich2012,Angelo2017,Esteves2017,Zhang2021,Deibert2021,Masson2024}. Additional transits and joint analyses with other facilities will be needed to improve constraints, as well as separate retrievals with non-\ce{H2}-dominated compositions to quantify their impact on posterior constraints and Bayesian model evidence.
  
  \subsection{AU Mic b}
  Discovered in 2020 \citep{Plavchan2020}, this warm sub-Neptune (0.037\,\Mjup, 0.37\,\Rjup, \citealp{Zicher2022}) orbits a young ($22\pm3$\,Myr, \citealp{Mamajek2014}) M-dwarf. The notorious host star activity has made previous atmospheric characterization attempts extremely challenging \citep[e.g.,][]{Palle2020_aumic,Rockcliffe2023,Masson2024}. With only a single transit available, incomplete removal of the stellar signature caused the \textit{Full Run} to converge toward stellar residuals rather than planetary signal. Retrieving meaningful constraints for this target required stronger model assumptions through the \textit{Upper Limits Run}, resulting in the abundance upper limits reported in Table~\ref{tab:upper_limits}. Our CCF analysis using these best-fit parameters revealed no planetary signature (Fig.~\ref{fig:55cnce_aumic_ccf}). Accumulating additional transit observations over longer timescales may eventually enable atmospheric detection by increasing signal-to-noise ratio while averaging out stellar activity contributions.

  \subsection{GJ 1214 b}
  Discovered in 2009 \citep{Charbonneau2009}, GJ\,1214\,b is a sub-Neptune (0.025\,\Mjup, 0.24\,\Rjup; \citealp{Cloutier2021}) on a 1.58-day orbit around an M dwarf. Analysis of the five transits from the \textit{Full Run} retrieval and CCF reveal strong low-Kp telluric residuals that prevent a robust atmospheric retrieval\footnote{Corner plots and CCF maps for GJ\,1214\,b, GJ\,436\,b, HAT-P-11\,b, and WASP-80\,b are accessible as Fig. 14 to 17 on \url{https://cloud.cab.inta-csic.es/s/a3G7DzExPtaF7L7}}. Letting only abundances free in the \textit{Upper Limits Run}, we obtain upper limits for all species except \ce{H2O}; the latter remains biased by telluric contamination (Table~\ref{tab:upper_limits}). This non-detection aligns with previously reported flat spectra and remains compatible with a dense high-altitude cloud deck and/or high metallicity \citep[e.g.,][]{Bean2010,Bean2011,Desert2011,Berta2012,deMooi2012,Miller2012_gj1214,Teske2013,Valencia2013,Kreidberg2014,Cloutier2021}.

  \subsection{GJ 3470 b}
  Discovered in 2012 \citep{Bonfils2012}, this sub-Neptune (0.04\,\Mjup, 0.35\,\Rjup; \citealp{Kosiarek2019}) orbits an M dwarf in 3.3 days and exhibits atmospheric escape \citep[e.g.,][]{Nina2020,Palle2020,Lampon2023,Allart2023,Masson2024}. Early observations yielded flat spectra, consistent with clouds, high metallicity, and \ce{CH4} depletion from disequilibrium chemistry \citep{Nascimbeni2013,Crossfield2013,Dragomir2015}. \ce{H2O} was detected with HST and Spitzer \citep{Benneke2019}, later confirmed by JWST \citep{Beatty2024} along with \ce{CH4}, \ce{CO}, \ce{CO2}, and \ce{SO2}. The \ce{SO2} detection further indicates that photochemical processes contribute to disequilibrium chemistry in GJ\,3470\,b's upper atmosphere. \citet{Dash2024} analyzed two CARMENES transits but were unable to detect \ce{H2O}. 
  
  Our \textit{Full Run} analysis (Fig.~\ref{fig:gj3470}) of five SPIRou transits yields puzzling constraints. The retrieved Kp (\red{$127\pm{13}$\,\kms}) is consistent with the planetary value, while a V$_0$ of \red{$-5.5\pm{0.6}$\,\kms} is plausible for atmospheric circulation. However, several parameters are physically problematic: the temperature (\red{$T_{\rm iso} = 1110^{+58}_{-74}$\,K}) is approximately twice the equilibrium value ($615\pm16$\,K; \citealp{Kosiarek2019}) and incompatible with \citet{Beatty2024}'s thermal profile. The retrieved \ce{CH4} abundance (\red{$\logt{\ce{CH4}}=-1.5^{+0.4}_{-0.5}$ MMR, or $-2.3^{+0.8}_{-0.7}$ Volume Mixing Ratio, VMR}) contradicts JWST's measurement ($-4.05^{+0.25}_{-0.27}$ VMR; \citealp{Beatty2024}). The OH abundance converges at the prior's upper edge, despite OH requiring ultrahot Jupiter conditions for \ce{H2O} dissociation \citep{Nugroho2021,Choi2025,Yang2025}. Furthermore, like \citet{Dash2024}, we detect no \ce{H2O}, \ce{CO}, or \ce{CO2}---the dominant JWST species \citep{Beatty2024}. These contradictions suggest residual stellar contamination biases the \textit{Full Run} results. The CCF computed from these abundances shows a \red{3.6}-$\sigma$ \ce{OH}-dominated peak near the expected (Kp, V$_0$) position. The corresponding RV shift is inconsistent with telluric contamination but could originate from M-dwarf activity \red{and uncorrected CLV or RME contributions}. However, the signal morphology is unusual: stellar contamination typically produces symmetric 'V-shaped' features extending to Kp $\sim 0$ \citep[see e.g.][]{Chiavassa2019}, absent in Figure~\ref{fig:gj3470}. One explanation could be that our reduction$-$notably the SysRem correction$-$preferentially subtracts stationary or slowly shifting stellar components, so the low-Kp stellar signal is corrected more strongly, while high-Kp stellar residual can persist and mimic a planetary. Alternatively, the presence of \ce{OH} in the upper atmosphere is required to explain the photochemically produced \ce{SO2} detected by \citet{Beatty2024}. It may therefore be possible to constrain an \ce{OH} abundance profile from their \ce{SO2} detection and estimate if the escaping planetary atmosphere could contribute to some extent to the observed CCF signal. Such analysis would require advanced photochemistry modeling coupled with atmospheric escape simulations and is therefore left for future work.

  \subsection{GJ 436 b}
  Discovered in 2004 \citep{Butler2004}, GJ\,436\,b is a sub-Neptune (0.45\,\Mjup, 0.43\,\Rjup; \citealp{Maxted2022}) on a 2.6-day orbit around an M dwarf. The planet hosts a large comet-like hydrogen exosphere \citep{Kulow2014,Ehrenreich2015,LaVie2017,DosSantos2019}, and Spitzer data indicate a CO-rich, \ce{CH4}-depleted, highly metallic atmosphere \citep{Stevenson2010,Madhusudhan2011,Line2014}. Previous ground-based efforts have reported non-detections of FeH \citep{Kesseli2020}, \ce{H2O}, \ce{CH4}, or \ce{CO} \citep{Grasser2024}. With only one SPIRou transit, the \textit{Full Run} did not yield meaningful constraints. Results from the \textit{Upper Limits Run} favor high \ce{CO2} and \ce{HCN} abundances, yet with no detection from the CCF analysis. Our search for atmospheric signature is therefore inconclusive and will require stacking additional transits, potentially with multiple instruments, to reach constraining sensitivity.

  \subsection{GJ 486 b}
  Discovered in 2021 \citep{Trifonov2021}, GJ\,486\,b is a 1.3\,R$_\oplus$, 2.8\,M$_\oplus$ planet on a 1.5-day orbit around an M dwarf. Formation scenarios span from a bare rock to an H/He envelope with solar or enhanced metallicity, with an H$_2$O-dominated atmosphere deemed most likely \citep{Caballero2022}. JWST observations reported a flat spectrum compatible with a water-rich atmosphere or an airless body \citep{Moran2023}. Ground-based searches reported non-detections of \ce{H2O}, \ce{CH4}, \ce{NH3}, \ce{HCN}, \ce{CO2}, CO \citep{Ridden2023} and no metastable He escape \citep{Masson2024}. Our \textit{Full Run} analysis of the five SPIRou transits shows no clear planetary signature, while the \textit{Upper Limits Run} statistically favors \ce{CO} (\red{$\logt{\ce{CO}} = -2.2^{+0.9}_{-1.0}$}) yet without supporting CCF detections (Fig.~\ref{fig:wasp69_gj486}). We therefore report no atmospheric detection, with upper limits indicated in Table~\ref{tab:upper_limits}.

  \subsection{HAT-P-11 b}
  Discovered in 2010 \citep{Bakos2010}, this low-mass hot-Neptune (25.7\,\Mearth, 4.7\,\Rearth) orbits a K star on a 4.89 days eccentric orbit ($e\sim0.26$; \citealp{Huber2017}), allowing for clear stellar-planetary separation. Prior studies detected \ce{H2O} \citep[e.g.,][]{Fraine2014,Cubillos2022} and atmospheric escape \citep[e.g.,][]{Allart2018,DosSantos2022,Masson2024}. \citet{Chachan2019} reported tentative \ce{CH4}, while \citet{Basilicata2024} detected \ce{H2O} and \ce{NH3}, with tentative \ce{CO2} and \ce{CH4}, but no \ce{C2H2}, \ce{HCN}, or \ce{H2S} from four GIANO-B transits. Our four-transit \textit{Full Run} yields no meaningful constraints. The \textit{Upper Limits Run} converges to extreme \ce{H2S} at prior boundaries and to \red{$\logt{\ce{CH4}}$ = $-1.3^{+0.7}_{-0.6}$} MMR, all without CCF confirmation. We conclude to a non-detection of HAT-P-11\,b atmosphere, and report 3-$\sigma$ upper limits in Table~\ref{tab:upper_limits}.

  \subsection{HD\,189733\,b} 
  Discovered in 2005 \citep{Bouchy2005}, HD\,189733\,b is a hot-Jupiter (1.16\,\Mjup, 1.12\,\Rjup; \citealp{Alonso2019_hd189733}) on a 2.2-day orbit around a K star. Its atmosphere exhibits well-established detections of \ce{H2O}, \ce{CO}, and \ce{CO2} from space and ground-based observations \citep[e.g.,][]{Swain2009,Birkby2013,Sing2016,Brogi2019,Alonso2019_hd189733,Boucher2021,Blain2024}, with recent JWST constraints on \ce{H2O}, \ce{CO2}, \ce{CO}, and \ce{H2S} \citep{Fu2024}. We analyzed eight SPIRou transits, excluding two with low SNR and poor time coverage (2020-07-03, 2020-07-05; Table~\ref{transit_tab}), and including the two transits studied by \citet{Boucher2021}. The \textit{Full Run} converges to \red{Kp = $136.3^{+5.3}_{-5.4}$\,\kms} and \red{V$_0$ = $-3.7\pm{0.2}$\,\kms} (Fig.~\ref{fig:hd189733_ccf}). This blueshift is consistent with prior measurements ($-3.9\pm1.3$\,\kms in \citealp{Alonso2019_hd189733}; $-4.62^{+0.46}_{-0.44}$\kms in \citealp{Boucher2021}) and theoretical predictions for day-night wind circulation in hot Jupiters. Rotational broadening \red{(V$_{\rm rot}$ = $1.4\pm{0.6}$\,\kms) is compatible within 2-$\sigma$ with the} expected equatorial velocity under tidal locking ($\sim2.6$\,\kms). Our best-fit atmospheric model disfavors high-altitude clouds (\red{log$_{10}P_{\rm cl}$ = $-1.9^{+0.5}_{-0.3}$\,bar}) and favors \red{$T_{\rm iso}$ = $830^{+100}_{-80}$}\,K, \ce{H2O} (\red{$-0.9\pm{0.2}$ MMR; $-1.7\pm{0.4}$ VMR}), and a high \ce{H2S} abundance although convergence to the upper edge of the prior prevented us from deriving a numerical constraint. Our $T_{\rm iso}$ is compatible within 2-$\sigma$ with \citet{Boucher2021} and \citet{Blain2024}. Our $\logt{\ce{H2O}}$ estimation is in agreement within 2-$\sigma$ with \citet{Blain2024} ($-1.50^{+0.12}_{-0.16}$ in MMR) but twice higher than \citet{Boucher2021} ($-4.4\pm0.4$ in VMR). Similarly, our $\logt{\ce{H2S}}$ value (\red{$-0.2\pm0.1$\,MMR; $-1.0\pm0.5$\,VMR}) is much higher than \citet{Fu2024} ($-4.5^{+0.57}_{-0.35}$ VMR). This discrepancy may result from our inclusion of more molecular species in the retrieval, resulting in a higher mean molecular weight, a more confined modeled atmosphere, and therefore increased abundances to reach similar line depths. The CCF analysis confirms \ce{H2O} at 5.0-$\sigma$ and tentatively detects \ce{H2S} at \red{3.4}-$\sigma$ with a Kp and V$_0$ offset toward lower values. Although not favored by the retrieval, we performed a CCF search for \ce{CO} using \citet{Fu2024}'s abundance and find a 4.4-$\sigma$ peak at Kp $\sim200$\,\kms and V$_0$ = $-17$\,\kms, which we suspect to be driven by uncorrected \red{CLV and RME} from the host star \citep{Brogi2019}. A combined \ce{H2O}+\ce{H2S} CCF model yields an unambiguous 5.3-$\sigma$ detection of the planet atmosphere. We therefore report presence of \ce{H2O}, and tentative detections of \ce{H2S} and \ce{CO}. The associated constraints are reported in Table~\ref{tab:retrieval_results}. 
  
  \subsection{HD 209458 b}
  Discovered in 1999 \citep{Henry2000}, HD\,209458\,b is a hot Jupiter (0.68\,\Mjup, 1.36\,\Rjup; \citet{Torres2008}) on a 3.5-day orbit around a G-type star. Prior detections include blueshifted \ce{CO} \citep{Snellen2010,Brogi2019,Giacobbe2021}, \ce{H2O} \citep{Sing2016,Tsiaras2016b,Tsiaras2018,Sanchez2019,Giacobbe2021,Xue2024}, and recent JWST observations reporting \ce{CO2} detection and absence of \ce{CH4}, \ce{C2H2}, and \ce{HCN} \citep{Xue2024}. Our single-transit study converged to a \ce{H2O} signature at \red{Kp = $121^{+25}_{-14}$}\,\kms and \red{V$_0$ = $-4.7^{+0.8}_{-0.7}$}\,\kms, with poorly constrained \Tiso and abundances. These Kp and V$_0$ values agree well with \citet{Sanchez2019} (Kp = $176^{+30}_{-38}$\,\kms; V$_0$ = $-5.2^{+2.6}_{-1.3}$\,\kms), yet our CCF search for \ce{H2O} only yield a 2.8-$\sigma$ local peak compatible with a planetary detection but too low even for a tentative detection (Fig.~\ref{fig:hd209}). The \textit{Upper Limits Run} mostly rules out $\logt{\ce{NH3}}$ above \red{$-3.9$}\,MMR (\red{-4.8 VMR;} Table.~\ref{tab:upper_limits}), \red{hence slightly improving \citet{Xue2024}'s constraints (\red{log(\ce{NH3})$\,\leq-4.2$\,VMR})}. We therefore conclude with a non detection of HD\,209458\,b atmosphere.

  \subsection{TOI-1807 b}
  \label{section_toi1807}
  Discovered in 2021 \citep{Hedges2021}, TOI-1807\,b is a young planet ($\sim300$\,Myr; $1.26\pm0.04$\,R$_\oplus$, $2.6\pm0.5$\,M$_\oplus$; \citealp{Hedges2021,Nardiello2022}) on a 0.55-day orbit around a K star. Previous He(2$^3S$) observations reported no atmospheric escape detection \citep{Gaidos2023,Orell2024,Masson2024}, though absence of He-escape does not rule out primary atmosphere presence, as non-He escape observations have been documented in similar targets with strong H-escape \citep{Munoz2025}. We performed here the first molecular signature search for TOI-1807\,b to our knowledge, using three SPIRou transits, and obtained puzzling results. The \textit{Full Run analysis} shows clear convergence at \red{Kp = $204^{+9}_{-12}$\,\kms compatible with} the expected value $\sim220$\,\kms, yet with a redshift of \red{V$_0 = 10.4^{+0.8}_{-1.4}$}\,\kms. The best model yields \red{$T_{\rm iso}=2840^{+760}_{-490}$}\,K, compatible with the $T_{\rm eq}=2100^{+39}_{-40}$ estimated by \citet{Hedges2021}, and an extremely high \ce{H2S} abundance (\red{$-0.4^{+0.1}_{-0.2}$ MMR}). Computing the CCF map including all species at their median posterior value yield a \red{4.4}-$\sigma$ tentative signal (Fig.~\ref{fig:toi1807}), with \ce{H2S} alone being tentatively detected at \red{4.2}-$\sigma$. As K stars are too hot to contain \ce{H2S}, the signal is not of stellar origin. However, the puzzling retrieved constraints call for further verification before claiming this signal as planetary. We performed a phase-dependent CCF analysis (Fig.~\ref{fig:toi1807_ccf_phase}), and obtained a tentative signature appearing only during transit. We further verified whether the signal would be consistent with the expected planetary RV trail using \red{the parameters reported by \citet{Nardiello2022} (see their Table 3) and the equation for keplerian semi-amplitude $K_p = \left( \frac{2\pi G}{P} \right)^{1/3} \frac{M_\star \sin i}{(M_\star + M_p)^{2/3}} \frac{1}{\sqrt{1 - e^2}}$. \red{The eccentricity being unconstrained, we allowed a value in the [0$-$0.1] range as is commonly observed for similarly young planets. Plotting the extremals planetary trails allowed by the currently reported uncertainties on the other parameters in the equation, we confirm that the observed K$_p$ and V$_0$ offsets could be explained by a current lack of precision in TOI-1807\,b orbital parameters (lower panel of Fig.~\ref{fig:toi1807_ccf_phase}). Additionally, 
  }
  the V$_0$ redshift may indicate further uncertainties in the transit ephemerides and potentially unconstrained transit-timing variations.} It is however difficult to trust this signal as planetary considering the extreme \ce{H2S} abundance and absence of other detected species. Detailed atmospheric escape and chemistry modeling combined with complementary observations are required to validate or refute this tentative signal, which, if confirmed, would be the first detection of TOI-1807\,b's atmosphere. As these considerations lies outside the scope of this study, we keep them as prospective for future work and report upper limits obtained following the \textit{Upper Limits Run} in Table~\ref{tab:upper_limits}.

  \subsection{WASP-127\,b}
  Discovered in 2016 \citep{Lam2017}, this hot-Jupiter (0.16\,\Mjup, 1.3\,\Rjup; \citealp{Stefansson2020}) orbits a G star on a 3.87-days retrograde, misaligned orbit \citep{Allart2020}. It's highly inflated atmosphere makes it an ideal target for atmospheric studies, with reported detections of \ce{H2O}, \ce{CO} and \ce{CO2} \citep{Welbanks2019,Spake2021,Boucher2023,Nortmann2025}. \citet{Nortmann2025} notably reported a double-peaked CCF consistent with a $7.7\pm0.2$\,\kms supersonic equatorial jet. Our \textit{Full Run} analysis of the five SPIRou transits (including three from \citealp{Boucher2023}) converges on an \ce{H2O} planetary signal at \red{Kp $=143.3^{+3.9}_{-3.5}$}\,\kms, \red{V$_0 = -6.6^{+0.3}_{-0.2}$}\,\kms, \red{$\logt{\ce{H2O}}=-0.6^{+0.2}_{-0.4}$} MMR (\red{$-1.4^{+0.5}_{-1.0}$} VMR), and \red{$T_{\rm iso}=740\pm{60}$}\,K (rounded to two significant digits from Fig.~\ref{fig:wasp127_ccf}). We note however that for \ce{H2O} convergence near the upper edge of the prior could potentially bias our posterior median and error estimates. The fit favors high-altitude clouds ({$\logt{P_{\rm cl}}=-4.9^{+0.4}_{-0.2}$}) and rotational broadening \red{V$_{\rm rot}=1.8\pm{0.5}$}\,\kms \red{compatible with tidally locked equatorial velocity} ($\sim1.15$\,\kms). The CCF analysis confirms our \ce{H2O} signal at \red{7.2}-$\sigma$ above CCF noise. High-altitude clouds are also favored in previous studies \citep{Skaf2020,Boucher2023}, and our results agree within 1-$\sigma$ confidence intervals with \citet{Boucher2023} ($V_0=-7.0\pm0.5$\,\kms, $\logt{\ce{H2O}}=-2.39^{+0.47}_{-0.63}$ in VMR, \Tiso$=769^{+180}_{-253}$) and the blueshifted signature in \citet{Nortmann2025}. We do not recover the tentative OH signal of \citet{Boucher2023}. Our data were not sensitive enough to detect the CO signature and $\pm8$\,\kms CCF-splitting reported by \citet{Nortmann2025}, though our $\logt{\ce{CO}}$ posterior (\red{$-1.6^{+0.8}_{-3.1}$} in MMR) would be compatible with their results. We report no detections for the other species and report our constraints in Table\,\ref{tab:retrieval_results}.
  
  \subsection{WASP-69\,b}
  Discovered in 2014, WASP-69\,b is a 0.26\,\Mjup, 1.05\,\Rjup hot Saturn on a 3.87\,days orbit around a K star \citep{Anderson2014}. Its large atmospheric scale height allowed for detection of atmospheric escape \citep[e.g.,][]{Nortmann2018,Wang2021,Allart2023,Masson2024} and molecules such as \ce{H2O}, \ce{CO}, and \ce{CO2} \citep{Tsiaras2018,Guilluy2022,Schlawin2024,Deka2025}. Detections of \ce{CH4}, \ce{NH3}, and \ce{C2H2} were reported by \citet{Guilluy2022} from transits observations with GIANO, while \citet{Schlawin2024} found no dayside \ce{CH4} from JWST observations. Analysing the single SPIRou transit, our \textit{Full Run} analysis converged on a low-Kp \ce{H2O} spurious feature that we attribute to telluric residuals. On the other hand, the \textit{Upper Limits Run} favors a mix of \ce{CO} (\red{$-1.0^{+0.5}_{-2.4}$} in $\mathrm{log_{10}}$-MMR), \ce{CH4} ({$-5.8^{+0.6}_{-0.9}$}), \ce{HCN} (\red{$-4.6^{+0.5}_{-0.6}$}), and \ce{C2H2} (\red{$-4.5^{+0.6}_{-0.7}$}), with convergence toward the upper edge of the prior for \ce{OH} ($-0.7^{+0.2}_{-0.4}$) (Fig.\,\ref{fig:wasp69_gj486}). Computing a CCF map based on this best model yields a 2.7-$\sigma$ local maximum compatible with the planetary Kp and $V_0=0$\,\kms yet with too low \snr for even a tentative detection. We therefore report no detection, with upper limits listed in Table~\ref{tab:upper_limits}. 

  \begin{table*}[]
  \centering
  \tiny
  \begin{threeparttable}
  \caption{\textit{Full Run} retrieval results.}
  \label{tab:retrieval_results}  
  \begingroup
  \setlength{\tabcolsep}{1.5pt}  
  \renewcommand{\arraystretch}{1.6}  
  
  {

  \begin{tabular}{@{}l|c|c|c|c|c|c|c|c|c|c|c|c|c|c@{}}
  \hline \hline
  Target & Kp & V$_0$ & T$_\mathrm{iso}$ & \multicolumn{9}{c|}{log$_{10}$-MMR} & \Vrot & log$_{10}$(P$_{\rm cloud}$) \\ 
  \noalign{\vskip -4pt}       
  & [\kms] & [\kms] & [K] & \ce{H2O} & \ce{CO} & \ce{CH4} & \ce{NH3} & \ce{CO2} & \ce{H2S} & \ce{HCN} & \ce{C2H2} & \ce{OH} & [\kms] & [bar] \\
  \noalign{\vskip -2pt}
  \hline
  HD 189733 b   & $136.3^{+5.3}_{-5.4}$ & $-3.7\pm{0.2}$ & $830^{+100}_{-80}$ & $-0.9\pm{0.2}$ & $\leq -0.9$ & $\leq -3.1$ & $\leq -3.0$ & $\leq -1.0$ & $-$ & $\leq -1.7$ & $\leq -1.5$ & $\leq -0.9$ & $1.4\pm{0.6}$ & $-1.9^{+0.5}_{-0.3}$ \\
  WASP-127 b    & $143.3^{+3.9}_{-3.5}$ & $-6.6^{+0.3}_{-0.2}$ & $740\pm{60}$ & $-0.6^{+0.2}_{-0.4}$ & $\leq -0.3$ & $\leq -2.0$ & $\leq -3.1$ & $\leq -0.9$ & $\leq -0.8$ & $\leq -1.7$ & $\leq -2.7$ & $\leq -1.1$ & $1.8\pm{0.5}$ & $-4.9^{+0.4}_{-0.2}$ \\
  WASP-76 b     & $221.3^{+6.1}_{-6.4}$ & $-8.9^{+1.0}_{-0.8}$ & $1550^{+240}_{-180}$ & $-2.2\pm{0.4}$ & $-0.4^{+0.1}_{-0.2}$ & $\leq -3.0$ & $\leq -3.4$ & $\leq -0.7$ & $\leq -1.0$ & $\leq -2.9$ & $\leq -2.3$ & $\leq -2.8$ & $6.2^{+1.0}_{-0.8}$ & $-2.7\pm{0.5}$ \\
  TOI-1807 b*   & $204^{+9}_{-12}$      & $-$                 & $2840^{+760}_{-490}$ & $\leq -3.7$ & $\leq -1.1$ & $\leq -3.3$ & $\leq -3.9$ & $\leq -1.7$ & $-0.4^{+0.1}_{-0.2}$ & $\leq -3.0$ & $\leq -2.1$ & $\leq -4.1$ & $\leq 11$ & $0.1^{+0.6}_{-0.7}$  \\
  GJ 3470 b*    & $127\pm{13}$          & $-5.5\pm0.6$        & $1110^{+58}_{-74}$ & $\leq -2.9$ & $\leq -0.7$ & $-1.5^{+0.4}_{-0.5}$ & $\leq -2.6$ & $\leq -1.3$ & $\leq -0.9$ & $\leq -2.2$ & $\leq -1.8$ & $-$ & $\leq 5.3$ & $-0.6 \pm {0.1}$ \\
  \hline
  \end{tabular}
  
  }

  \endgroup
  \begin{tablenotes}[flushleft]
  \footnotesize
  \item \textbf{Notes.} Results obtained from the \textit{Full Run} analysis. Values are given as median posteriors with 1-$\sigma$ uncertainties. Upper limits are given at the 3-$\sigma$ level. Errors where further pessimistically rounded to a maximum of 2 digits. Targets marked with an * correspond to tentative atmospheric detections.
  \end{tablenotes}
  \end{threeparttable}
  \end{table*}

  \begin{table}[]
  \centering
  \tiny
  \begin{threeparttable}
  \caption{\textit{Upper Limits Run} results.}
  \label{tab:upper_limits}  
  \begingroup
  \setlength{\tabcolsep}{1.5pt}  
  \renewcommand{\arraystretch}{1.6} 

  {

  \begin{tabularx}{0.48\textwidth}{@{}l|
  >{\centering\arraybackslash}X|
  >{\centering\arraybackslash}X|
  >{\centering\arraybackslash}X|
  >{\centering\arraybackslash}X|
  >{\centering\arraybackslash}X|
  >{\centering\arraybackslash}X|
  >{\centering\arraybackslash}X|
  >{\centering\arraybackslash}X|
  >{\centering\arraybackslash}X@{}}
  \hline \hline     
  Target & \ce{H2O} & \ce{CO} & \ce{CH4} & \ce{NH3} & \ce{CO2} & \ce{H2S} & \ce{HCN} & \ce{C2H2} & \ce{OH} \\
  \noalign{\vskip -2pt}
  \hline
  55 CnC e      & $-3.7$ & $-0.2$ & $-1.5$ & $-0.5$ & $-$ & $-1.0$ & $-$ & $-1.1$ & $-0.5$ \\ 
  AU Mic b      & $-2.4$ & $-0.5$ & $-4.3$ & $-4.2$ & $-0.2$ & $-1.9$ & $-1.5$ & $-2.9$ & $-3.0$ \\ 
  GJ 1214 b     & $-$ & $-0.7$ & $-5.3$ & $-6.0$ & $-3.5$ & $-4.6$ & $-3.9$ & $-3.9$ & $-1.8$ \\ 
  GJ 3470 b     & $-4.3$ & $-0.2$ & $-4.5$ & $-6.1$ & $-3.5$ & $-3.6$ & $-5.4$ & $-5.0$ & $-$ \\ 
  GJ 436 b      & $-0.4$ & $-0.2$ & $-0.3$ & $-0.6$ & $-$ & $-0.8$ & $-$ & $-$ & $-0.2$ \\ 
  GJ 486 b      & $-3.8$ & $-0.3$ & $-3.7$ & $-5.3$ & $-1.5$ & $-1.8$ & $-4.4$ & $-4.5$ & $-1.1$ \\ 
  HAT-P-11 b    & $-1.2$ & $-$ & $-$ & $-3.9$ & $-0.1$ & $-$ & $-1.3$ & $-1.1$ & $-1.0$ \\ 
  HD 209458 b   & $-1.1$ & $-$ & $-$ & $-3.9$ & $-0.8$ & $-0.8$ & $-1.0$ & $-1.1$ & $-0.3$ \\
  TOI-1807 b    & $-2.7$ & $-$ & $-2.1$ & $-0.5$ & $-0.1$ & $-$ & $-0.2$ & $-1.8$ & $-1.6$ \\ 
  WASP-69 b     & $-3.7$ & $-0.1$ & $-3.4$ & $-5.5$ & $-3.2$ & $-4.4$ & $-3.2$ & $-2.8$ & $-0.1$ \\ 
  WASP-80 b     & $-$ & $-1.3$ & $-3.5$ & $-4.3$ & $-0.4$ & $-0.6$ & $-1.1$ & $-0.4$ & $-0.1$ \\ 
  \hline
  \end{tabularx}
  
  }

  \endgroup
  \begin{tablenotes}[flushleft]
  \footnotesize
  \item \textbf{Notes.} 3-$\sigma$ upper limits obtained from the \textit{Upper Limits Run} (Sec.\,\ref{NS_method}). The priors on $\rm log_{10}$(MMR) were set to be uniform over the [-8 $-$ 0] range (Tab.\,\ref{tab:priors_run1}).
  \end{tablenotes}
  \end{threeparttable}
  \end{table}

  \subsection{WASP-76\,b} 
  Discovered in 2016 \citep{West2016}, this ultrahot Jupiter (0.9\,\Mjup, 1.85\,\Rjup; \citealp{Ehrenreich2020}) orbits a F star in 1.8 days. Previous studies reported potential atmospheric escape \citep{Casasayas-Barris2021,Lampon2023,Masson2024} and molecular detections of \ce{H2O}, \ce{CO}, \ce{OH}, \ce{HCN}, \ce{TiO}, and \ce{VO} \citep[e.g.,]{Tsiaras2018,Landman2021,Fu2021,Sanchez2022,Yan2023,Maguire2024,Hood2024,Gandhi2024}. \citet{Ehrenreich2020} notably reported asymmetric CCF-detection of Fe attributed to nightside iron-clouds condensation. \citet{Hood2024}'s SPIRou retrieval disfavors clouds above $10^{-4}$\,bar and resulted in \ce{H2O} and \ce{CO} detections at different Kp values which they attribute to atmospheric circulation. We analyzed three SPIRou transits (including the 2020-10-31 observation of \citealp{Hood2024}). The \textit{Full Run} shows clear convergence to \ce{H2O} and \ce{CO} at \red{Kp $=221.3^{+6.1}_{-6.4}$}\,\kms and \red{V$_0 = -8.9^{+1.0}_{-0.8}$}\,\kms, with \red{$\logt{\ce{H2O}}=-2.2\pm0.4$} MMR (\red{$-3.1\pm0.6$} VMR) and \red{$\logt{\ce{CO}}=-0.4^{+0.1}_{-0.2}$} MMR (\red{$-1.3^{+0.4}_{-0.7}$} VMR) (Fig.~\ref{fig:wasp76_ccf}). The best model yields \red{$T_{\rm iso}=1550^{+240}_{-180}$}\,K, rotational broadening \red{V$_{\rm rot}=6.2^{+1.0}_{-0.8}$}\,\kms \red{compatible with tidally locked rotation ($\sim 5.2$\,\kms)}, and high-altitude clouds at \red{log$_{10}P_{\rm cl}=-2.7\pm{0.5}$}\,bar. From the CCF analysis, the \ce{H2O} signal is confirmed at 5.6-$\sigma$ while the \ce{CO} signature is tentatively detected at \red{3.0}-$\sigma$. Our abundances are consistent with \citet{Sanchez2022} and slightly exceed \citet{Hood2024}'s cloud-free results, which is expected given our model's preference for high-altitude clouds and therefore the need for higher abundances to fit the same signal. We estimated the C/O ratio to be nearly solar (C/O$ \,\sim1$), but uncertainties in the \ce{H2O} and \ce{CO} abundances prevent us from concluding toward a super- or sub-solar C/O value. \red{We furthermore note that our free chemistry setup cannot reliably constrain C/O in ultrahot Jupiters given the expected non-constant water MMR vertical profile due to dissociation. Further constraining the C/O ratio will therefore require a complementary dedicated retrieval accounting for chemistry and photodissociation.} Other species were not detected, and we report our constraints in Table\,\ref{tab:retrieval_results}.

  \begin{figure}[!htbp]
    \centering
    \includegraphics[width=0.88\columnwidth,trim=0.4cm 0.58cm 0cm 0.3cm,clip]{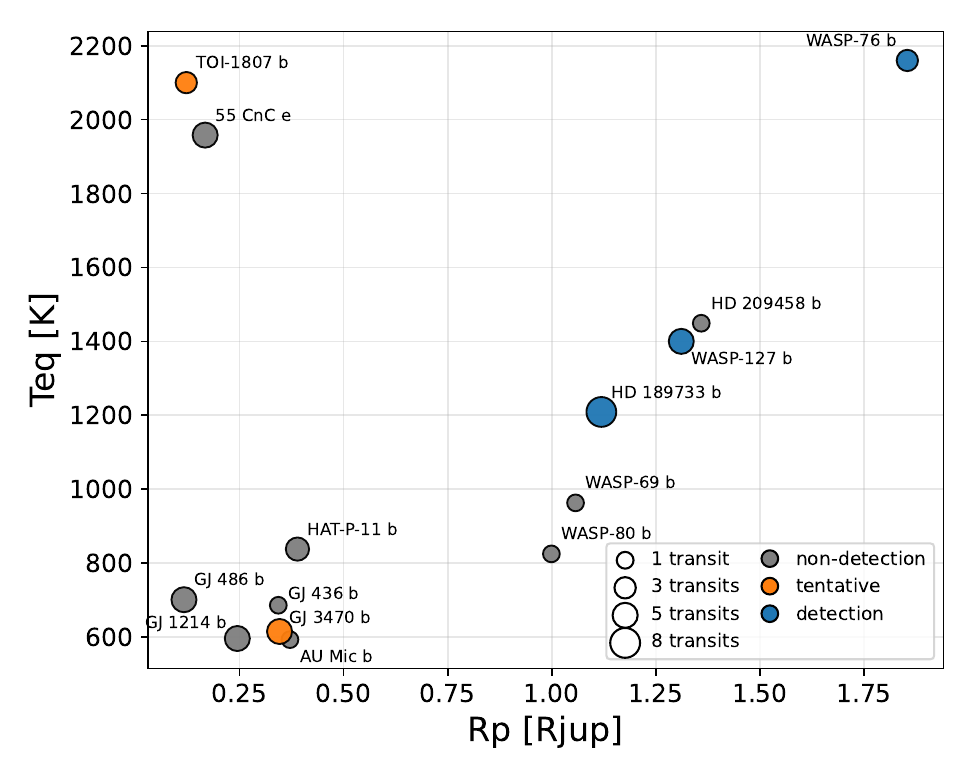}
    \caption{Radius$-$equilibrium temperature summary diagram of the 14 targets. Marker size scales with the number of analyzed transits, and colors indicate whether an atmosphere was detected (blue), tentatively detected (orange), and non-detected (gray).}
    \label{fig:Teq_rp_summary}
  \end{figure}

  \begin{figure*}[!htbp]
    \centering
    \includegraphics[trim={0cm 0cm 0cm 0cm}, clip, width=\textwidth]{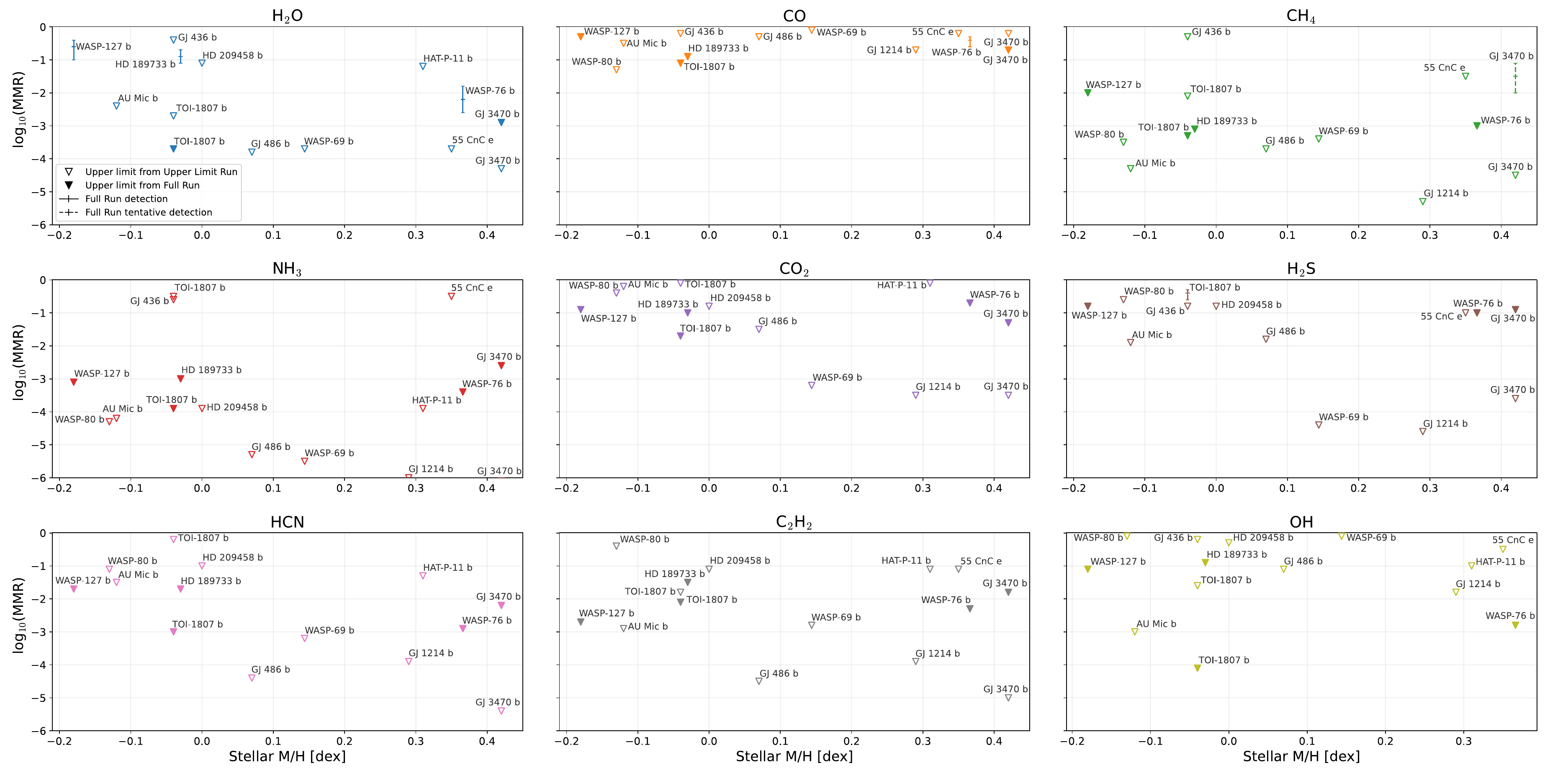}
    \caption{Constraints obtained on log$_{10}$(MMR) for each target as a function of host-star metallicity (log scale, relative to solar). Upper limits are shown as open and filled triangles for the \textit{Upper Limits Run} and \textit{Full Run}, respectively. Constraints from the \textit{Full Run} are shown with solid and dashed error bars for robust and tentative atmospheric detections, respectively. Error bars and upper limits correspond to the 1-$\sigma$ and 3-$\sigma$ uncertainties reported in Tables~\ref{tab:retrieval_results} and~\ref{tab:upper_limits}. Note that TOI-1807\,b and GJ\,3470\,b display two sets of constraints in some subplots: one from the \textit{Full Run}, and one from the \textit{Upper Limits Run}.}
    \label{fig:molecular_detec_summary}
  \end{figure*}

  \subsection{WASP-80\,b}
  WASP-80\,b (0.54\,\Mjup, $\sim1$\,\Rjup; \citealp{Triaud2013,Triaud2015}) is a warm Jupiter on a 3-day orbit around a relatively active K star. No atmospheric escape has been detected \citep{Fossati2022,Fossati2023,Allart2023,Masson2024}, despite predictions for strong escape \citep{Salz2016_a}. Prior work reported detections of \ce{H2O}, \ce{CO}, \ce{CH4}, \ce{NH3}, \ce{HCN}, and tentative \ce{CO2} \citep{Carleo2022,Bell2023,Jacobs2023}. Our single SPIRou transit lacks the sensitivity to independently confirm these detections. The \textit{Full Run} converges on a spurious telluric-driven \ce{H2O} signature, while the \textit{Upper Limits Run} yields unrealistically high \ce{H2O} abundances (attributed to telluric residuals) and marginal \ce{C2H2} at \red{$\logt{\ce{C2H2}}=-2.1^{+0.6}_{-0.8}$}, unconfirmed by CCF analysis. We therefore report no atmospheric detection and provide upper limits in Table~\ref{tab:upper_limits}. 

  \subsection{Results overview}
  Our results from the \textit{Full Run} and \textit{Upper Limits Run} are compiled in Tables\,\ref{tab:retrieval_results} and\,\ref{tab:upper_limits}, respectively. For visualization, we provide a summary figure showing the atmospheric detection status of each target as a function of the number of combined transits in the radius--equilibrium temperature plane (Fig.\,\ref{fig:Teq_rp_summary}). Unambiguous atmospheric detections are achieved only for hot-Jupiter targets with at least three combined transits, highlighting the need to stack additional observations to reach detection thresholds for other systems. To search for population trends, we derived the ranges of C/O, \ce{H2O}/H, and atmospheric metallicity permitted by the constraints obtained for each target in both runs. Plotting these quantities as a function of host-star metallicity or equilibrium temperature does not reveal a clear population-level trend, mainly because most upper limits remain weakly constraining. Reaching a level of precision suitable for robust population studies will require additional observations and complementary wavelength coverage. The derived MMR constraints per target as a function of host-star metallicity are shown in Fig.\,\ref{fig:molecular_detec_summary}, while additional figures are available as supplementary material in the associated repository\footnote{See Fig. 18 to 20 on \url{https://cloud.cab.inta-csic.es/s/a3G7DzExPtaF7L7}}. Although no clear correlation is inferred, these figures provide a useful overview of the sensitivity of our retrieval framework to each molecule: \ce{NH3} is generally the most strongly constrained species (i.e., with the lowest upper limits), likely because of its dense line forest in the SPIRou wavelength range and its negligible contribution in Earth's atmosphere and M-dwarf spectra. By contrast, \ce{H2O}, \ce{CO}, \ce{CO2}, and \ce{OH} are the most challenging species to constrain, likely because of stronger telluric contamination and, for M-dwarf hosts, residual stellar contamination.

  \section{Summary and Conclusions}
  \label{conclusion}
  We conducted a global analysis of 50 observations of 14 transiting exoplanets obtained with SPIRou at CFHT, combining public PI-archival data with data from three large programs (ATMOSPHERIX, SLS, SPICE). We developed an automated pipeline for the homogeneous reduction and analysis of large ground-based high-resolution spectroscopy datasets for atmospheric characterization, including a novel, unbiased automated optimization of SysRem or PCA for correlated residuals correction.

  We performed a blind-search for nine molecules across all targets, using Nested Sampling retrievals cross-validated with Cross Correlation Function (CCF) analyses. We detect \ce{H2O} and a tentative \ce{CO} signal in HD\,189733\,b, \ce{H2O} in WASP-127\,b, and \ce{H2O} with a tentative \ce{CO} signal in WASP-76\,b, in agreement with previous studies. These robust detections allowed us to constrain orbital radial velocity, atmospheric temperature, cloud pressure deck, and atmospheric circulation. While these constraints rely on strong model assumptions due to computational limitations, their consistency with prior work validates our automated blind-search approach for the remainder of the dataset.

  We also report several unexpected, potentially spurious, tentative signals that warrant further investigation. We find a tentative \red{3.4}-$\sigma$ \ce{H2S} signature in HD\,189733\,b which may support recent JWST observations. For GJ\,3470\,b, we identify a tentative \ce{OH} signal at \red{3.6}-$\sigma$, although its planetary origin cannot be unambiguously distinguished from residual stellar contamination from the M-dwarf host. Finally, we detect a tentative \red{4.4}-$\sigma$ signal in TOI-1807\,b using a model including all species at their best posterior values, with \ce{H2S} alone producing a tentative \red{4.2}-$\sigma$ signal. The location of the CCF peak in ($K_\mathrm{p}$, $V_0$) space is compatible with a planetary origin given current orbital uncertainties and, if confirmed, could indicate a slight eccentricity. However, assessing the physical plausibility of the inferred atmospheric properties will require further theoretical exploration beyond the scope of this work. At the population scale, our comparative analysis does not reveal any clear trend, mainly due to weakly informative upper limits. Additional observations, combined with complementary wavelength coverage, will be necessary to obtain constraints suitable for population-level analyses.

  The pipeline and methodology presented here are scalable to larger planetary samples and readily adaptable to other ground-based high-resolution spectrographs. As ground-based high-resolution spectroscopy provides a critical complement to space-based observations, efficient and automated reduction and analysis frameworks are becoming essential to keep pace with the increasing scale and ambition of atmospheric characterization efforts. Continued development is underway to improve telluric and stellar correction, reduce computational costs, and enable more complex atmospheric models in preparation for upcoming facilities and missions such as Ariel and ANDES.

  \section*{Acknowledgments}
   We thank the anonymous referee for their constructive comments and suggestions, which helped improve the quality and clarity of this manuscript. This research has been funded by the grants nº CNS2023-144309 by the Spain Ministry of Science, Innovation and Universities (MICIU/AEI/10.13039/501100011033) and NextGenerationEU/PRTR grants CNS2023-144309 and PID2023-150468NB-I00. This work is supported by the French Research National Agency (ANR) through the RaD3-net project (ANR-21-CE49-0020). S. Vinatier acknowledges funding from the Action Thématique ExoSystèmes and Programme National de Planétologie of CNRS/INSU and from the CNES. F. Debras acknowledges fundings from the French National Research Agency (ANR) project ExoATMO (ANR-25-CE49-6598), the \textit{Action Thématique Physique Stellaire} (ATPS), and the \textit{Action Thématique ExoSystèmes} (AT-EXOS) of CNRS/INSU co-funded by CEA and CNES. R. Allart acknowledges the Swiss National Science Foundation (SNSF) support under the Post-Doc Mobility grant P500PT\_222212 and the support of the Institut Trottier de Recherche sur les Exoplanètes (IREx). V. Yariv acknowledges funding from the physics doctoral school (ED-PHYS) of the Grenoble-Alpes University. This work is based on observations obtained at the Canada-France-Hawaii Telescope (CFHT) which is operated from the summit of Maunakea by the National Research Council of Canada, the Institut National des Sciences de l’Univers of the Centre National de la Recherche Scientifique of France, and the University of Hawaii. The observations at the Canada-France-Hawaii Telescope were performed with care and respect from the summit of Maunakea which is a significant cultural and historic site. This work has made use of the NASA Exoplanet Archive (\href{https://exoplanetarchive.ipac.caltech.edu/}{exoplanetarchive.ipac.caltech.edu}), The Extrasolar Planets Encyclopaedia (\href{exoplanet.eu}{exoplanet.eu}), the Astrophysics Data System (\href{https://ui.adsabs.harvard.edu/}{ui.adsabs.harvard.edu}), and the SIMBAD database (\href{https://simbad.u-strasbg.fr/simbad/}{simbad.u-strasbg.fr}). This work has made use of the following Python packages: \texttt{Numpy} \citep{harris2020array}; \texttt{SciPy} \citep{2020SciPy-NMeth}; \texttt{Astropy} \citep{astropy:2013, astropy:2018, astropy:2022}; \texttt{PyAstronomy} \citep{pya}; \texttt{statsmodel} \citep{Seabold2010}; \texttt{Jupyter} \citep{jupyter}; \texttt{Matplotlib} \citep{Hunter:2007}; \texttt{PyMultinest} \citep{Buchner2014}; \texttt{batman} \citep{Kreidberg2015}; \texttt{petitRADTRANS} \citep{Molliere2019}; and \texttt{Exo-REM} \citep{Baudino2015,Baudino2017,Charnay2018,Blain2021}.


\bibliographystyle{aa}
\bibliography{references}


\begin{appendix}

\section{Sigma clipping}
\label{std_sigma_clipping}
Once cleaned with SysRem, we apply a last sigma clipping step to remove any remaining outliers that may have survived or arised during the previous reduction steps (e.g., extreme values resulting from the division by an almost null value). For this step, our rejection criterion is based on the spectral distribution of the individual spectral bins' error $\sigma_{\lambda_i}$, approximated as the standard deviation of each spectral bin $\lambda_i$ along the time axis. How much the $\{\sigma_{\lambda_i}\}_{i=1}^{N_{\lambda}}$ (with $N_{\lambda}$ the number of spectral bins) are dispersed around the true $\sigma(\lambda)$ statistic is determined by the standard error $SE_{\sigma}$, whose expression can be approximated as (see, e.g., eq.~3 in \citealp{Ahn2003}):

\begin{equation}
  \label{SE}
  {SE}_\sigma = \frac{\sigma(\lambda)}{\sqrt{2(N_{obs}-1)}}
\end{equation}

With $\mathrm{N}_{obs}$ the number of observed exposures. Since we do not have access to the true $\sigma(\lambda)$ statistic, we approximated it by fitting the $\{\sigma_{\lambda_i}\}$ distribution with a LOWESS\footnote{We used the implementation from the python module \texttt{statsmodels}.} function \citep[locally weighted scatterplot smoothing;][]{Cleveland1979}, which uses local linear regression to fit a smooth curve to a set of data points with low sensitivity to outliers. We empirically set the rolling fitting window to comprise 20\% of the data, and found this method to yield a better fit than classical polynomial models. We further added an Earth atmospheric transmission correction term T($\lambda$) to account for the fact that spectral bins lying in low-transmission ranges would exhibit a higher noise level without necessarily being outliers. Replacing $\sigma(\lambda)$ by it's approximated values from the LOWESS fit $L(\lambda)$ in eq.\,\ref{SE}, our final rejection criterion for each spectral bin $\lambda_i$ can be written as:

\begin{align}
  \centering
  |\sigma_{\lambda_i} - \frac{L(\lambda_i)}{\sqrt{T(\lambda_i)}}| > 3\,\frac{L(\lambda_i)}{\sqrt{2(N_{obs}-1)}}
\end{align}

where we reject any spectral bin $\lambda_i$ whose standard deviation along time departs from more than 3-$\mathrm{SE}_\sigma$ from the LOWESS fit $L$, iterating until no more values are rejected with a maximum of 5 iterations. 
\begin{figure}[!htbp]
  \centering
  \includegraphics[trim={0cm 0cm 0cm 0cm}, clip, width=\hsize]{"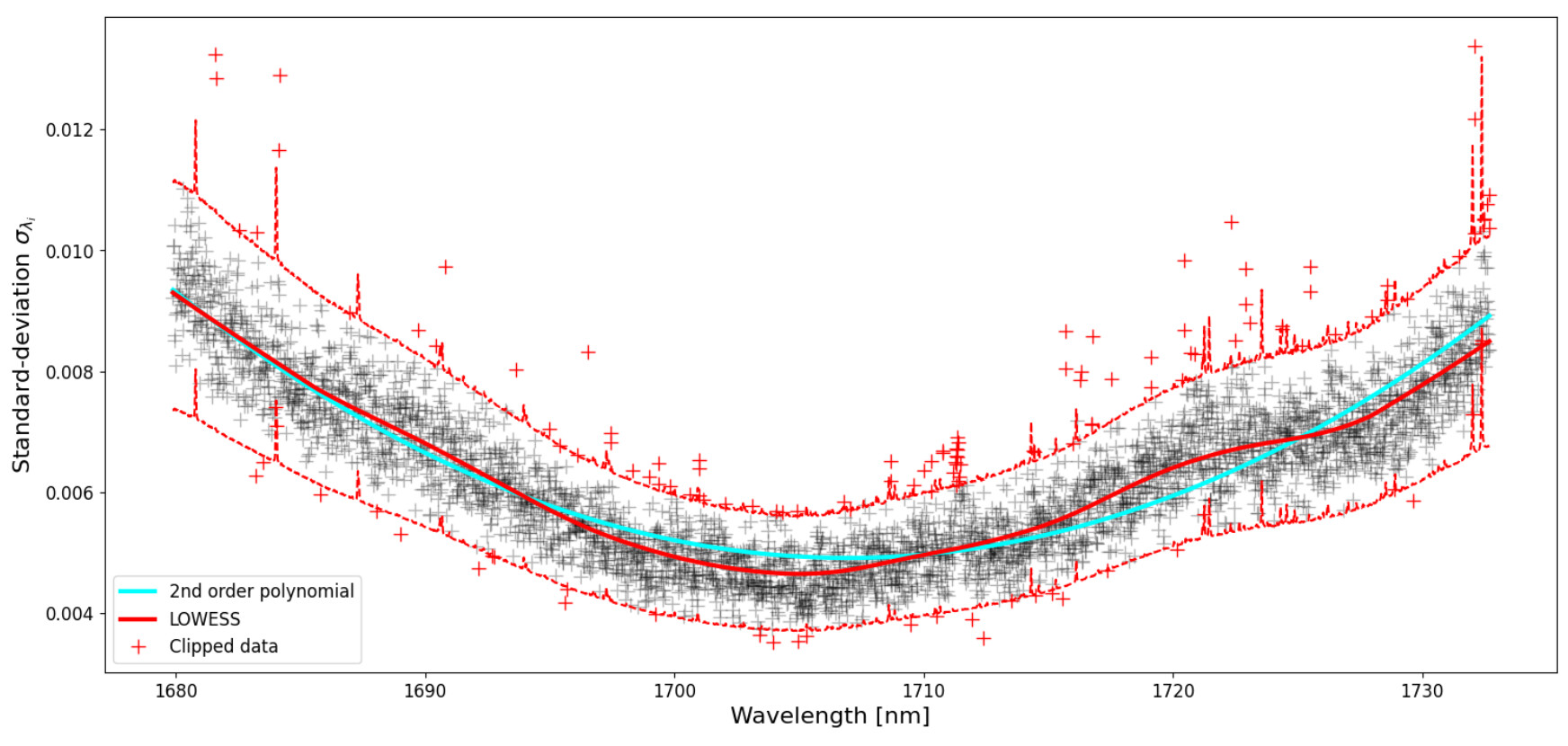"}
  \caption[Sigma clipping along the spectral axis]{Standard deviation computed for each spectral bin $\lambda_i$ along the time axis as a function of the wavelength. A single SPIRou spectral order is shown here for illustration. The LOWESS fit (red) yield a better approximation of the error trend than the second-order polynomial fit (cyan), as for higher-degree polynomial fits (not shown). The upper and lower rejection thresholds are shown in red dashed lines, with the apparent peaks corresponding to telluric absorption lines. Data rejected after the first sigma clipping iteration are shown in red crosses.}
\end{figure}

\vfill
\pagebreak

\section{Additional tables}
\begin{table}[h]
  \centering
  \caption{Linelists references for each species used in \texttt{petitRADTRANS}.}
  \label{tab:opacity_sources}
  \begin{tabular}{ll}
  \hline
  \hline
  Species & Reference \\
  \hline
  \ce{H2O} & HITEMP \citep{Rothman2010} \\
  \ce{CO} & HITEMP \citep{Rothman2010} \\
  \ce{CH4} & HITEMP \citep{Hargreaves2020} \\
  \ce{NH3} & HITRAN \citep{Rothman2013} \\
  \ce{CO2} & HITEMP \citep{Rothman2010} \\
  \ce{H2S} & HITRAN \citep{Rothman2013} \\
  \ce{HCN} & \citet{Harris2006} \\
  \ce{C2H2} & HITRAN \citep{Rothman2013} \\
  \ce{OH} & HITRAN \citep{Rothman2013} \\
  \hline
  \end{tabular}
\end{table}
  
\noindent

\begin{table}[h]
  \centering
  \caption{Prior distributions used in the Nested Sampling analysis for all targets. All distributions are uniform within the specified range. The log$_{10}$-MMR prior refer to the molecular abundances of all included species (\ce{H2O}, \ce{CO}, \ce{CH4}, \ce{NH3}, \ce{CO2}, \ce{H2S}, \ce{HCN}, \ce{C2H2}, and \ce{OH}).}
  \label{tab:priors_run1}
  \begin{tabular}{ll}
  \hline\hline
  Parameter & Prior (Run 1) \\
  \hline
  K$_p$ [\kms] & $\mathcal{U}$(0, 500) \\
  V$_0$ [\kms] & $\mathcal{U}$(-20, 5) \\
  \Tiso\ [$10^3$ K] & $\mathcal{U}$(0.2, 5) \\
  log$_{10}$(P$_{\rm cloud}$) [bar] & $\mathcal{U}$(-6, 1) \\
  log$_{10}$-MMR & $\mathcal{U}$(-8, 0) \\
  \Vrot\ [\kms] & $\mathcal{U}$(0, 15) \\
  \hline
  \end{tabular}
\end{table}

\section{Additional Corner Plots and CCF maps}

\begin{figure}[h]
    \centering
        
    \includegraphics[width=0.49\textwidth]{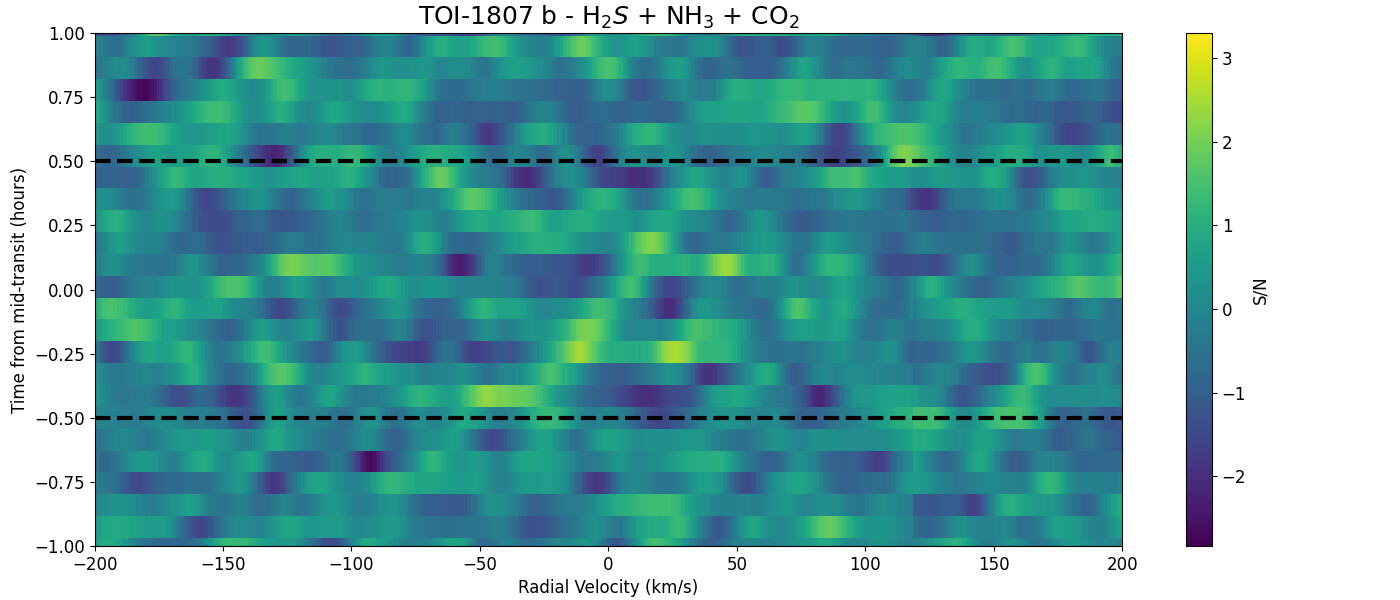}
        
    \includegraphics[width=0.49\textwidth]{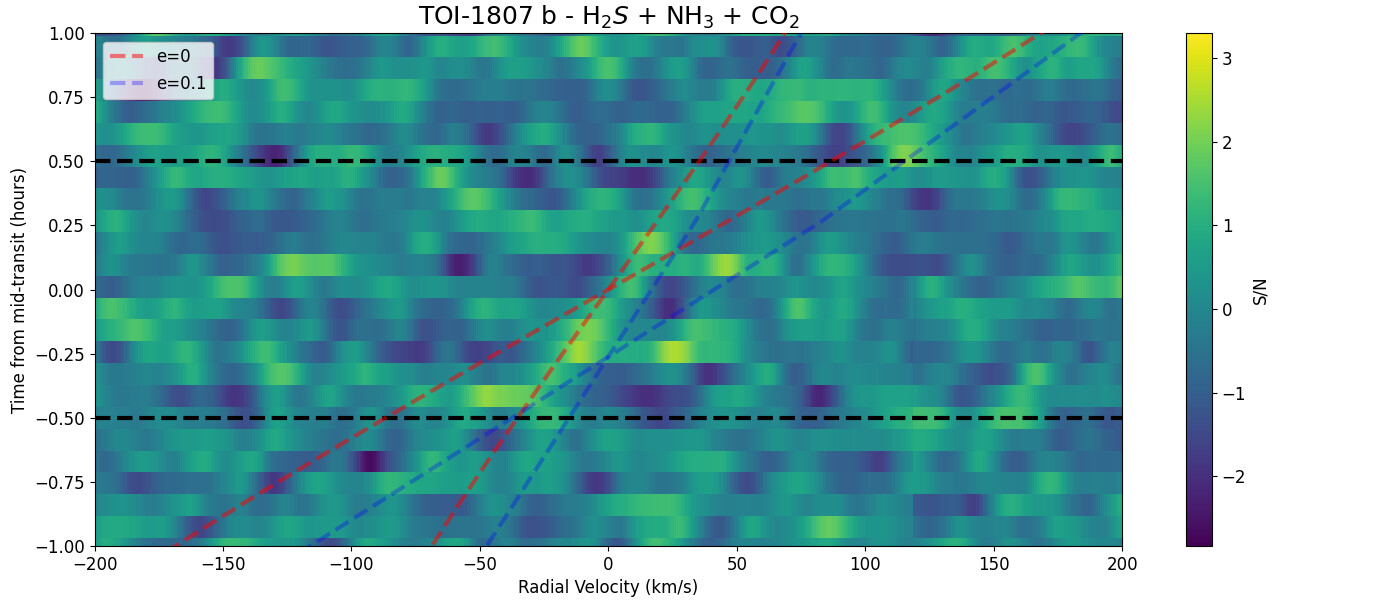}
        
    \caption{\textbf{Top:} TOI-1807 phase-resolved CCF in the stellar rest frame as a function of planet's radial velocity and time from mid-transit. Horizontal black lines indicate start and end of transit; color indicates the S/N. \textbf{Bottom:} Same figure showing the planetary radial-velocity trail region permitted by uncertainties in M$_p$, M$_\star$, \red{Period P, and inclination i (see the K$_P$ equation in Section~\ref{section_toi1807}), for zero (red) and 0.1 (blue) eccentricity.}}
    \label{fig:toi1807_ccf_phase}
\end{figure}

\begin{figure*}[!htbp]
  \centering
  \includegraphics[width=\textwidth]{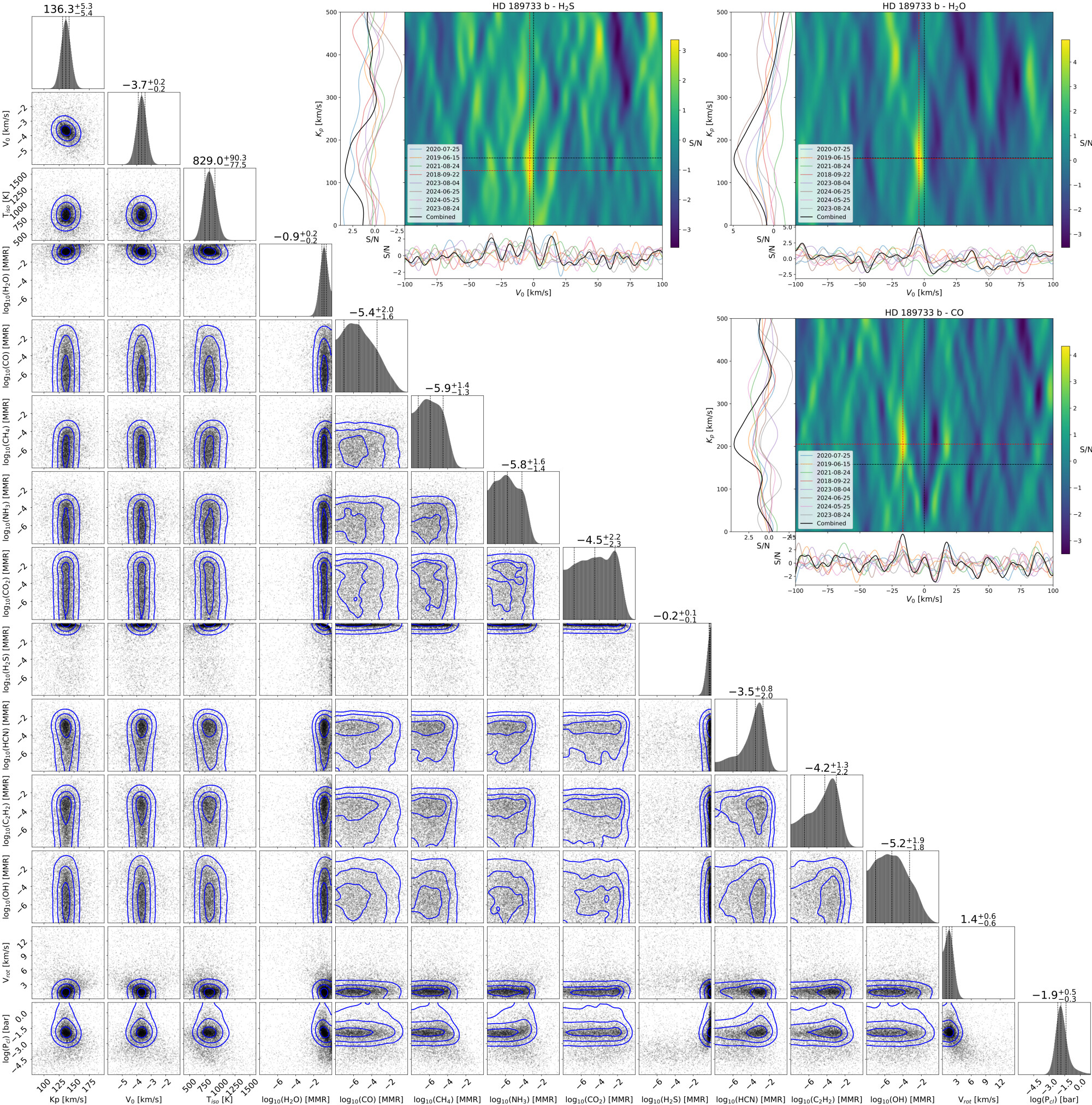}
  
  \caption{Same as Fig.~\ref{fig:wasp127_ccf} for HD\,189733\,b. \red{The CCF maps were obtained using a model built following the Section~\ref{NS_method} setup, including only \ce{H2O} or CO, with all parameters (\Tiso, MMR, \Vrot, $\log \rm P_{cloud}$) set to their median posterior values (indicated above each posterior distribution in the corner plot).}}
  \label{fig:hd189733_ccf}
\end{figure*}

\begin{figure*}[!htbp]
  \centering
  \includegraphics[width=\textwidth]{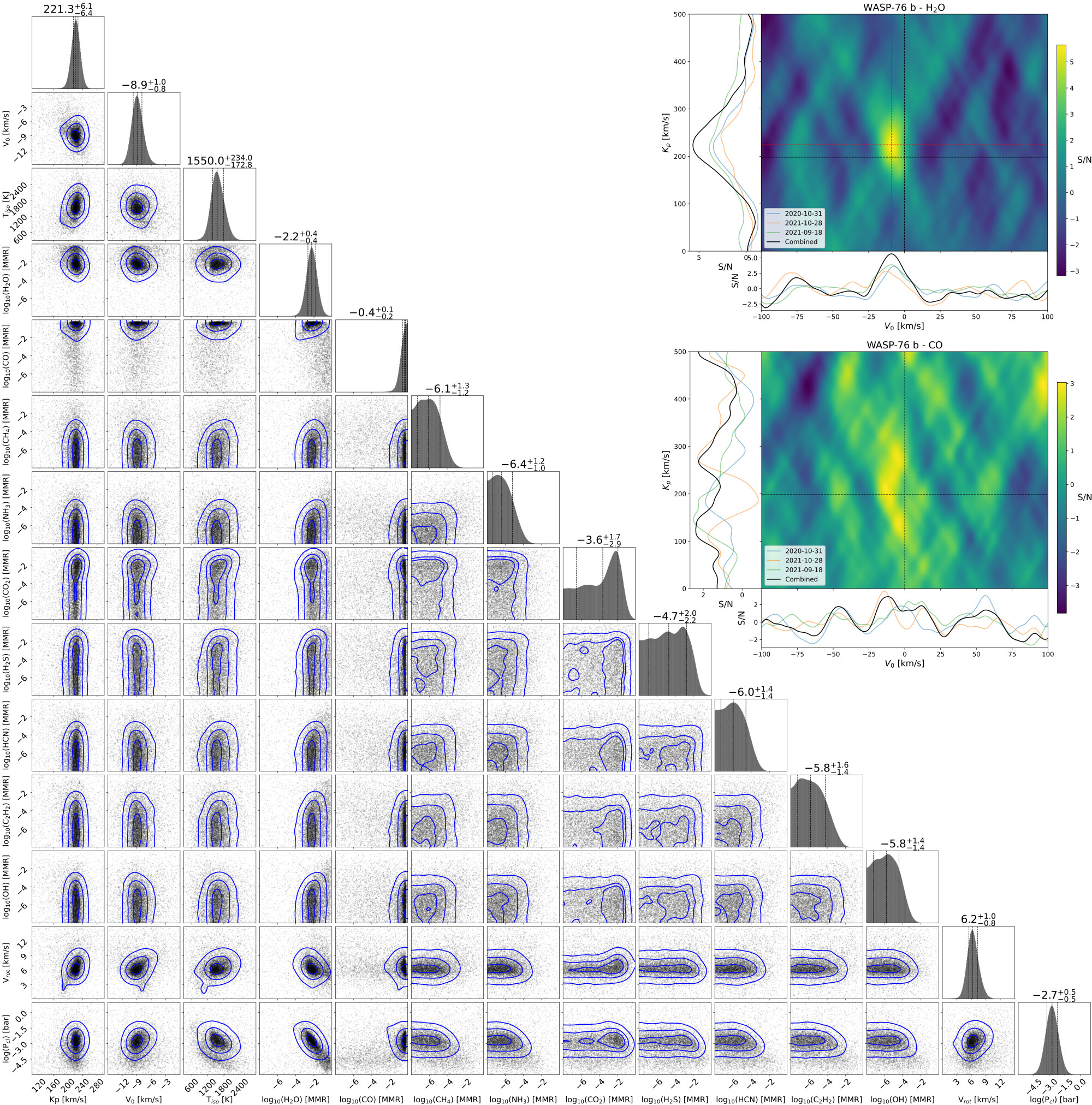}
  \caption{Same as Fig.~\ref{fig:wasp127_ccf} for WASP-76\,b. \red{The CCF maps were obtained using a model built following the Section~\ref{NS_method} setup, including only \ce{H2O} or CO, with all parameters (\Tiso, MMR, \Vrot, $\log \rm P_{cloud}$) set to their median posterior values (indicated above each posterior distribution in the corner plot).}}
  \label{fig:wasp76_ccf}
\end{figure*}

\begin{sidewaysfigure*}
  \centering
  \includegraphics[width=0.495\linewidth,trim=0cm 0cm 0cm 0cm,clip]{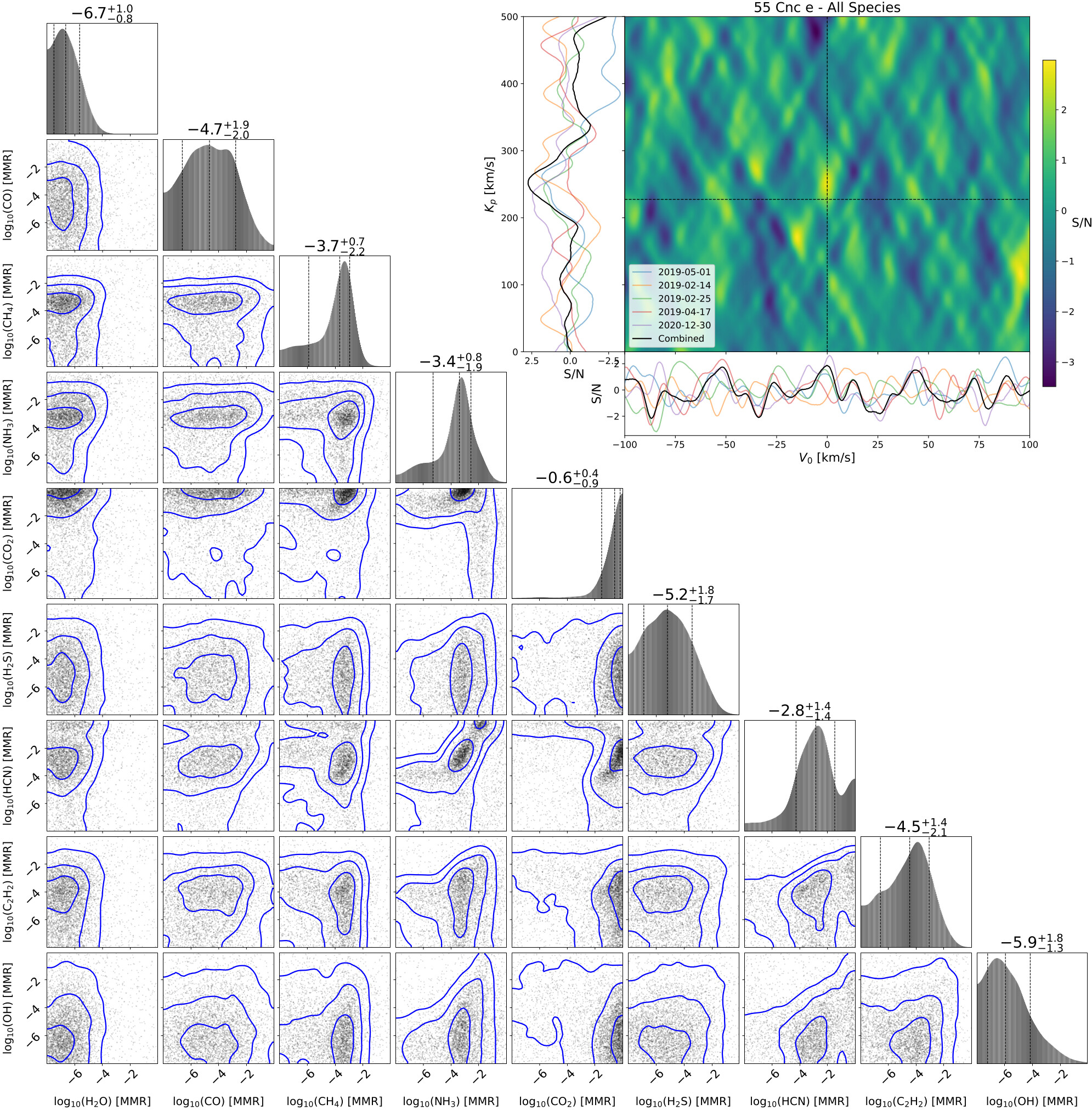}\hfill
  \includegraphics[width=0.495\linewidth,trim=0cm 0cm 0cm 0cm,clip]{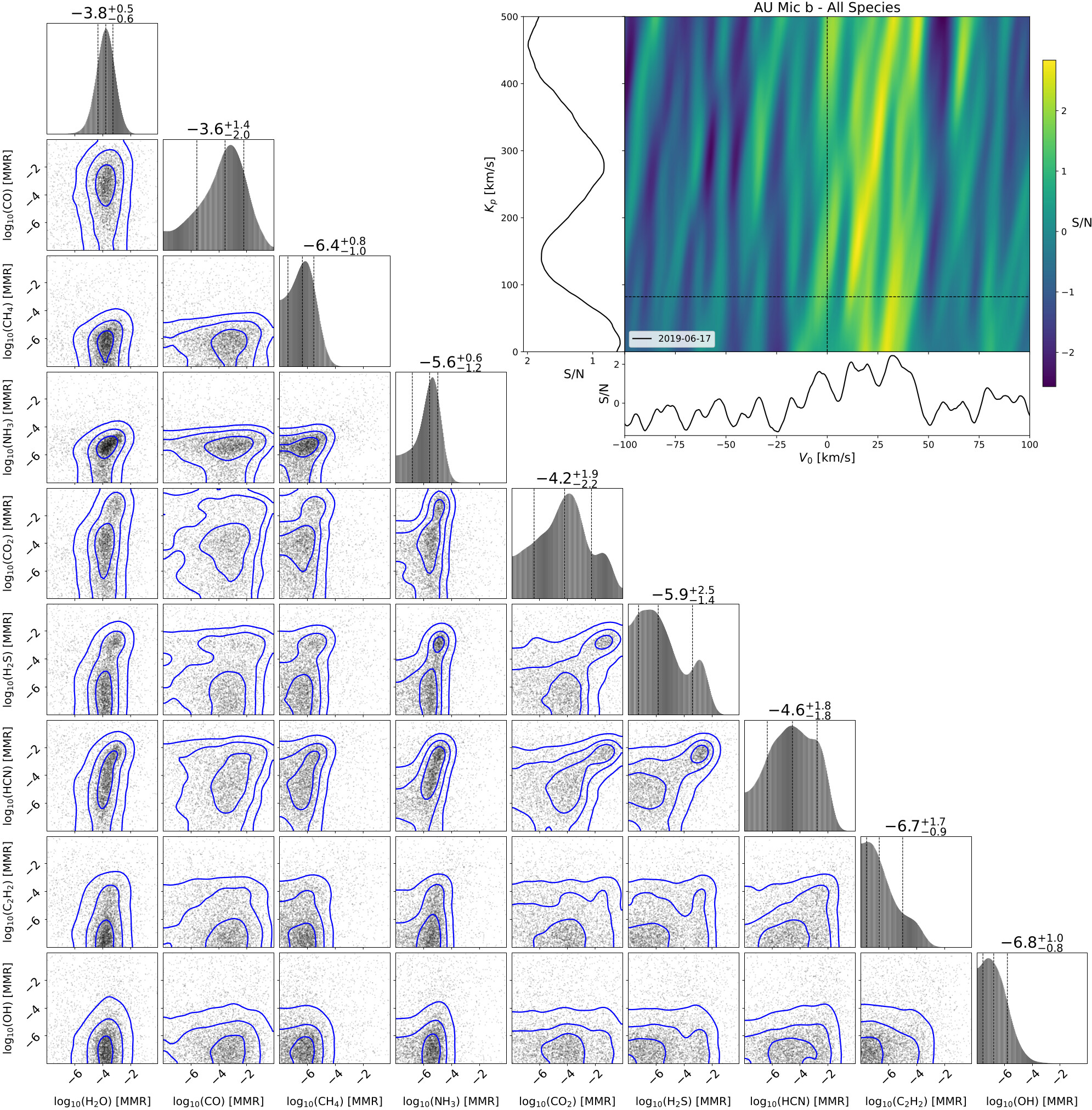}
  \caption{Corner plot from the \textit{Upper Limits Run} and associated CCF map for 55\,Cnc\,e (left) and AU\,Mic\,b (right). The posterior distributions were obtained by fixing the thermal profile to Exo-REM's predictions with a cloud-free model. The CCF maps were obtained with an atmospheric model including all species \red{with abundances set at their median posterior value (indicated above each posterior distribution in the corner plot) and using the Exo-REM's thermal profile}. The projected CCF at the expected planetary Kp and V$_0=0$\,\kms is shown in the left and lower CCF figure's panels for each transit date and for their combination (solid black). \textbf{Left:} the posteriors converged toward an unrealistically high \ce{CO2} abundance for 55\,Cnc\,e, with no detection in the CCF albeit for an inconclusive local $\sim3\sigma$ maxima at the planetary signature's expected position (black cross). \textbf{Right:} no planetary signature was found in the AU\,Mic\,b CCF map, the local peak near V$_0\sim10$\,\kms and Kp $\sim200$\kms being associated to remaining stellar residuals.}
  \label{fig:55cnce_aumic_ccf}
\end{sidewaysfigure*}

\begin{sidewaysfigure*}
  \centering
  \includegraphics[width=0.495\linewidth,trim=0cm 0cm 0cm 0cm,clip]{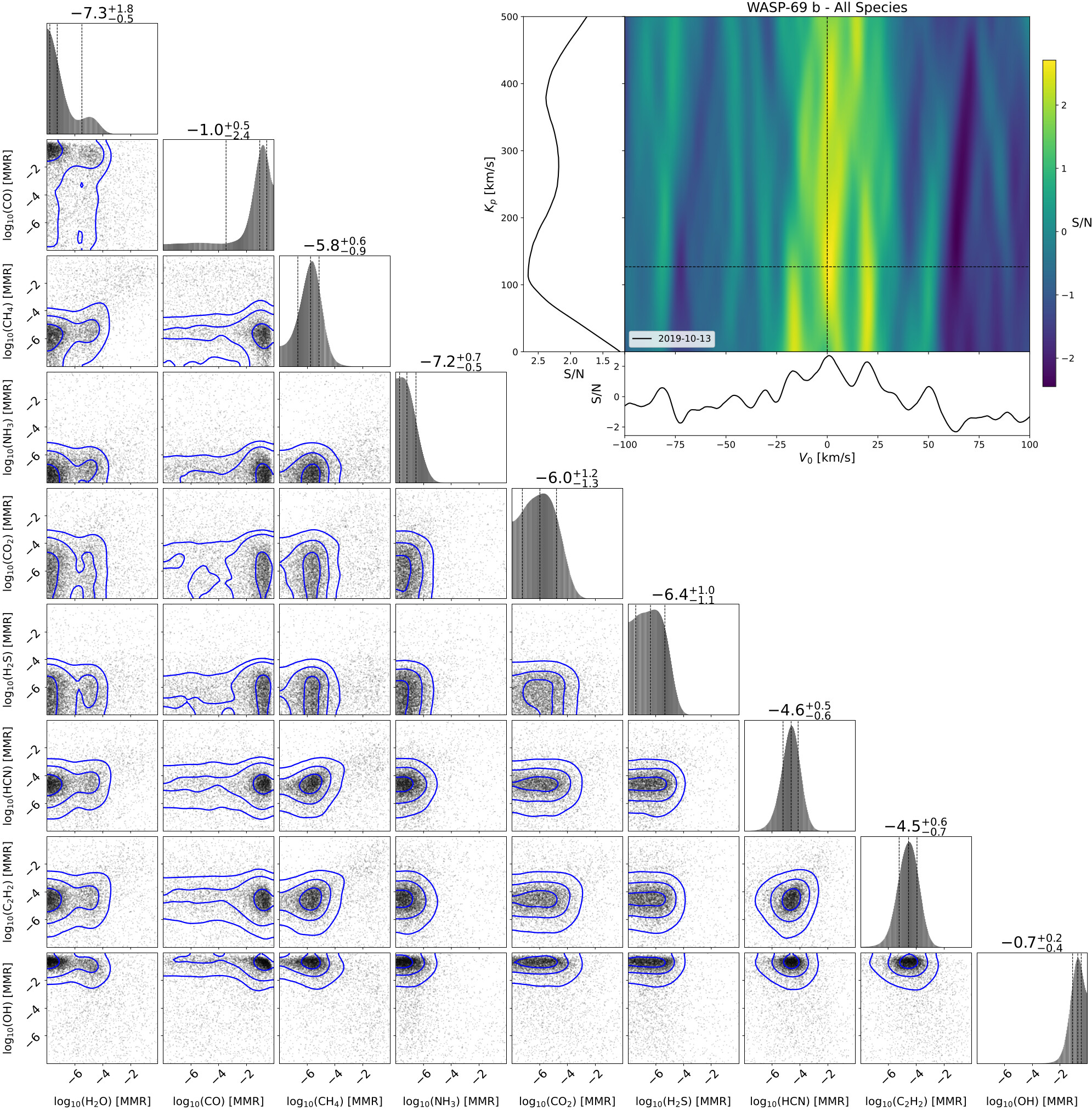}\hfill
  \includegraphics[width=0.495\linewidth,trim=0cm 0cm 0cm 0cm,clip]{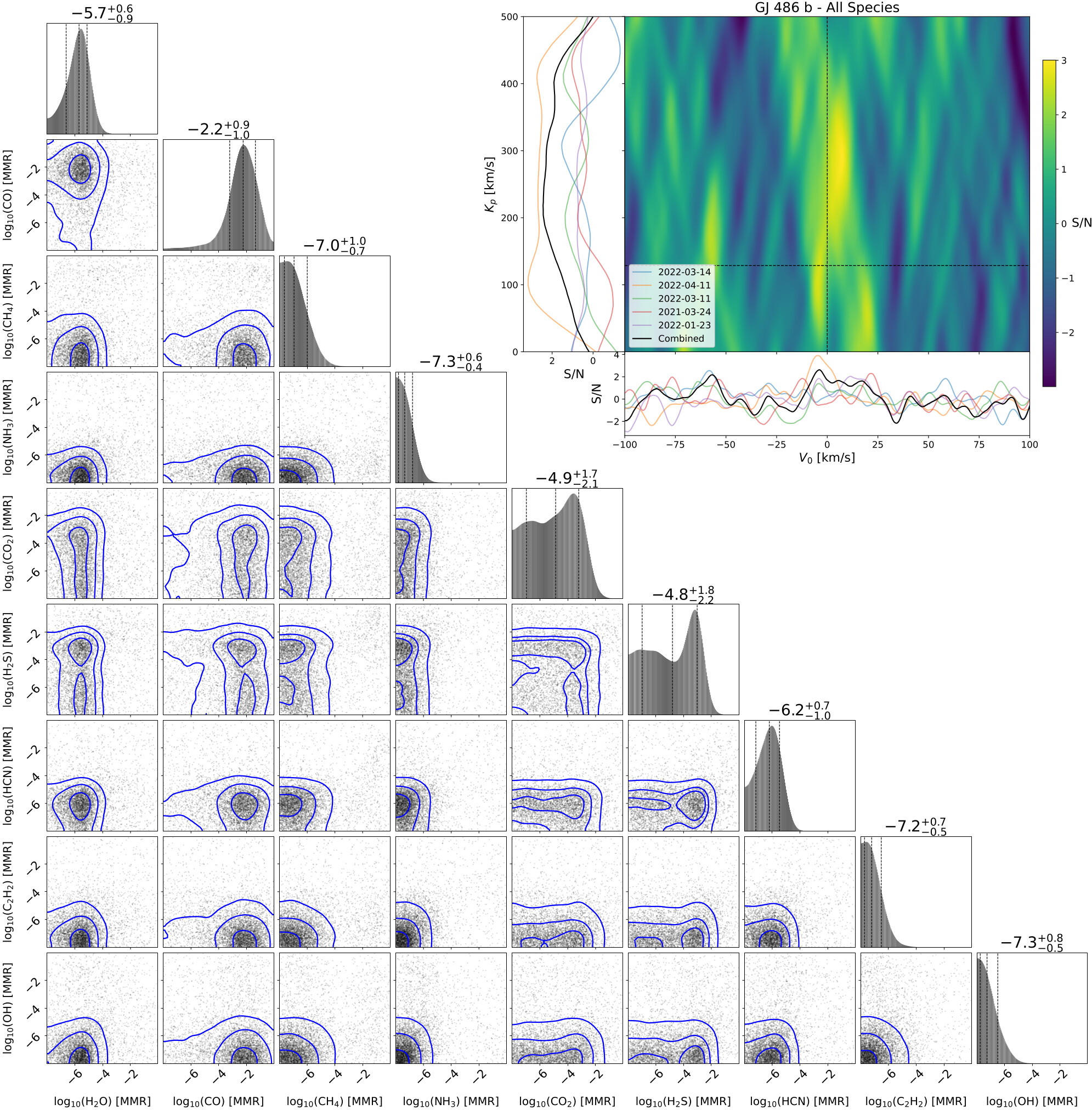}
  
  \caption{Corner plot from the \textit{Upper Limits Run} and associated CCF map for WASP-69 b (left) and GJ\,486\,b (right). The blue contours represent the 1-, 2-, and 3-$\sigma$ confidence intervals. The posterior distributions were obtained by fixing the thermal profile to Exo-REM's predictions with a cloud-free model. The CCF maps were obtained with an atmospheric model including all species \red{with abundances set at their median posterior value (indicated above each posterior distribution in the corner plot) and using the Exo-REM's thermal profile}. The projected CCF at the expected planetary Kp and V$_0=0$\,\kms (black cross) is shown in the left and lower CCF figure's panels for each transit date and for their combination (solid black). \textbf{Left:} although no convergence was found in the \textit{Full run}, the posteriors from the \textit{Upper Limit Run} displayed here seems to converge for \ce{CO}, \ce{CH4}, \ce{HCN}, \ce{C2H2}, and at the upper edge of the prior for \ce{OH}. The CCF map computed with all these species at their median posterior values exhibits in a 2.7-$\sigma$ signal at the expected planetary position, though too weak to report a detection even tentative of the planet atmosphere. \textbf{Right:} the high \ce{CO} abundance favored for GJ\,486\,b may result from residual signatures of the M-dwarf host, and the CCF map obtained with all species at their best abundance shows no sign of a planetary signature.}
  \label{fig:wasp69_gj486}
\end{sidewaysfigure*}

\begin{sidewaysfigure*}
  \centering
  \includegraphics[width=0.595\linewidth,trim=0cm 0cm 0cm 0cm,clip]{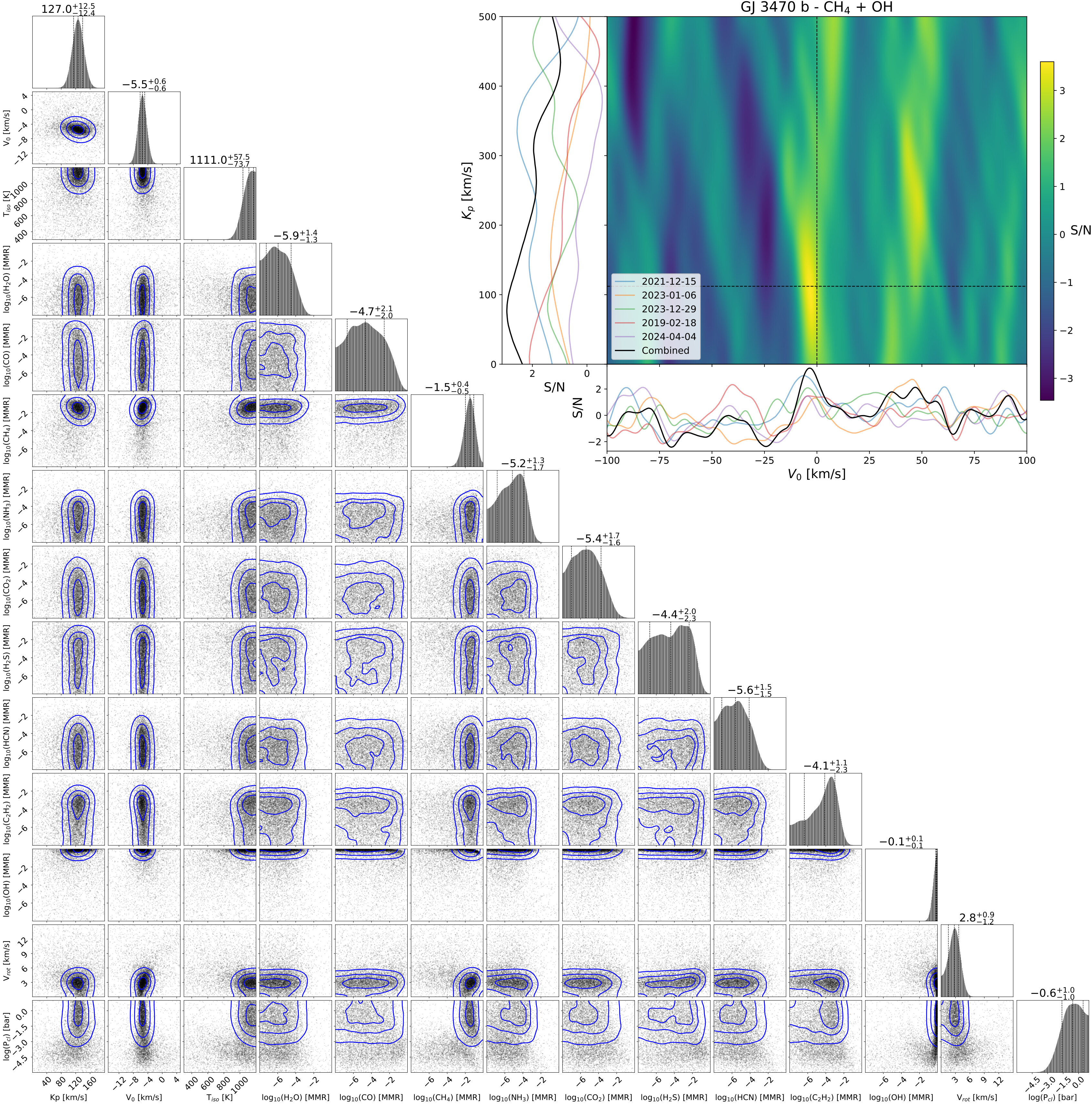}\hfill
  \includegraphics[width=0.395\linewidth,trim=0cm 0cm 0cm 0cm,clip]{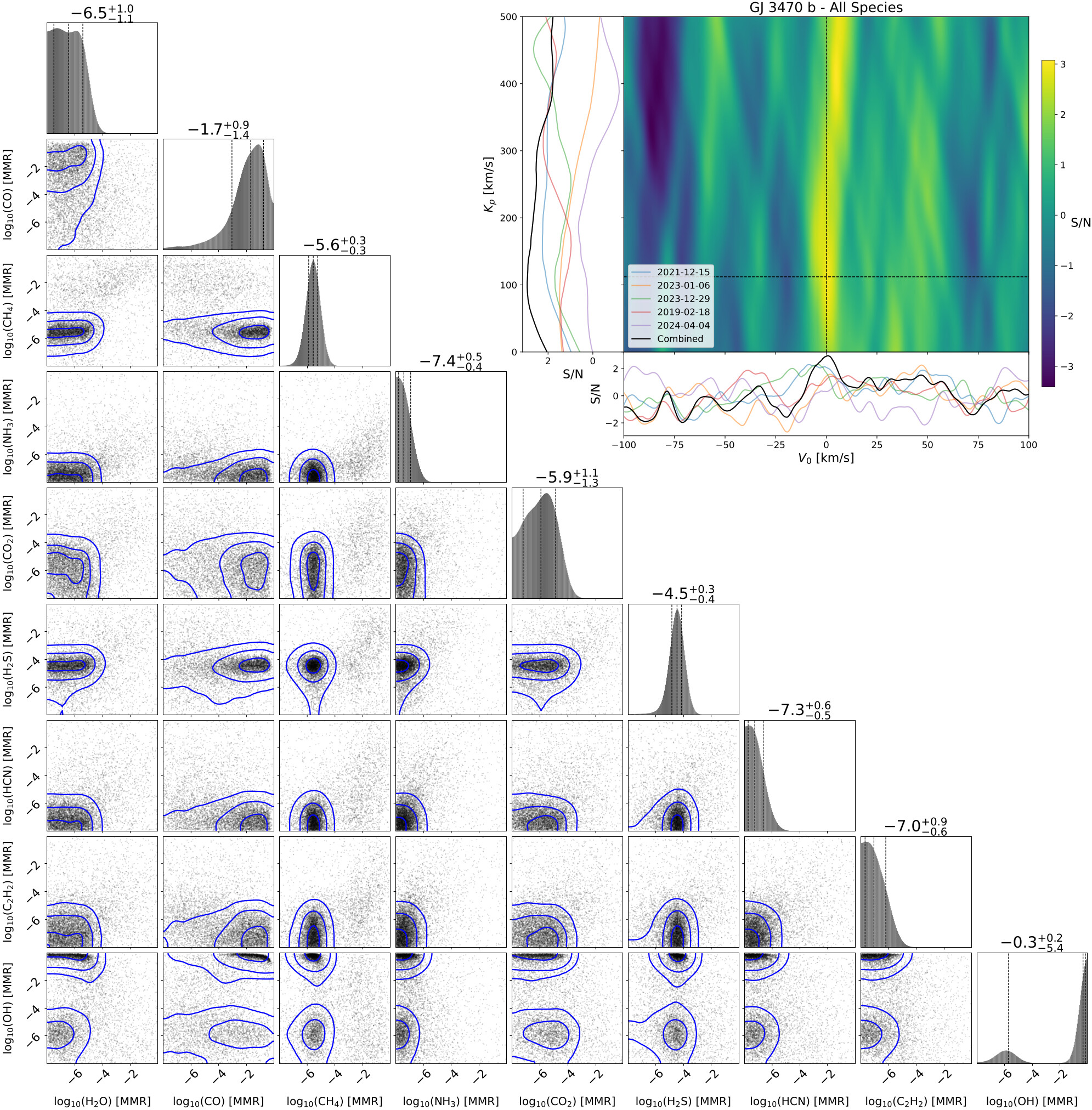}
  \caption{Corner plot from the \textit{Full} (left) and \textit{Upper Limits} (right) Runs with associated CCF maps for GJ\,3470\,b. The blue contours represent the 1-, 2-, and 3-$\sigma$ confidence intervals. \textbf{Left:} the posteriors converged toward a signature compatible with the planetary signal (black cross in the CCF), corresponding to a \red{3.6}-$\sigma$ tentative signature when computing a CCF map with an OH + \ce{CH4} model with \Tiso, \Vrot and log(P$_{cl}$) fixed to their median posterior values. The projected CCF at the expected planetary Kp and V$_0=0$\,\kms (black cross) is shown in the left and lower CCF figure's panels for each transit date and for their combination (solid black). \textbf{Right:} posteriors and CCF map obtained from the \textit{Upper Limits Run}, i.e. by fixing the thermal profile to Exo-REM's predictions with a cloud-free model. The CCF map was obtained using the best abundances, including all species \red{and using the Exo-REM's thermal profile}, with the lower and left panels showing the CCF projected along the theoretical planet's position (black cross).}
  \label{fig:gj3470}
\end{sidewaysfigure*}

\begin{sidewaysfigure*}
  \centering
  \includegraphics[width=0.595\linewidth,trim=0cm 0cm 0cm 0cm,clip]{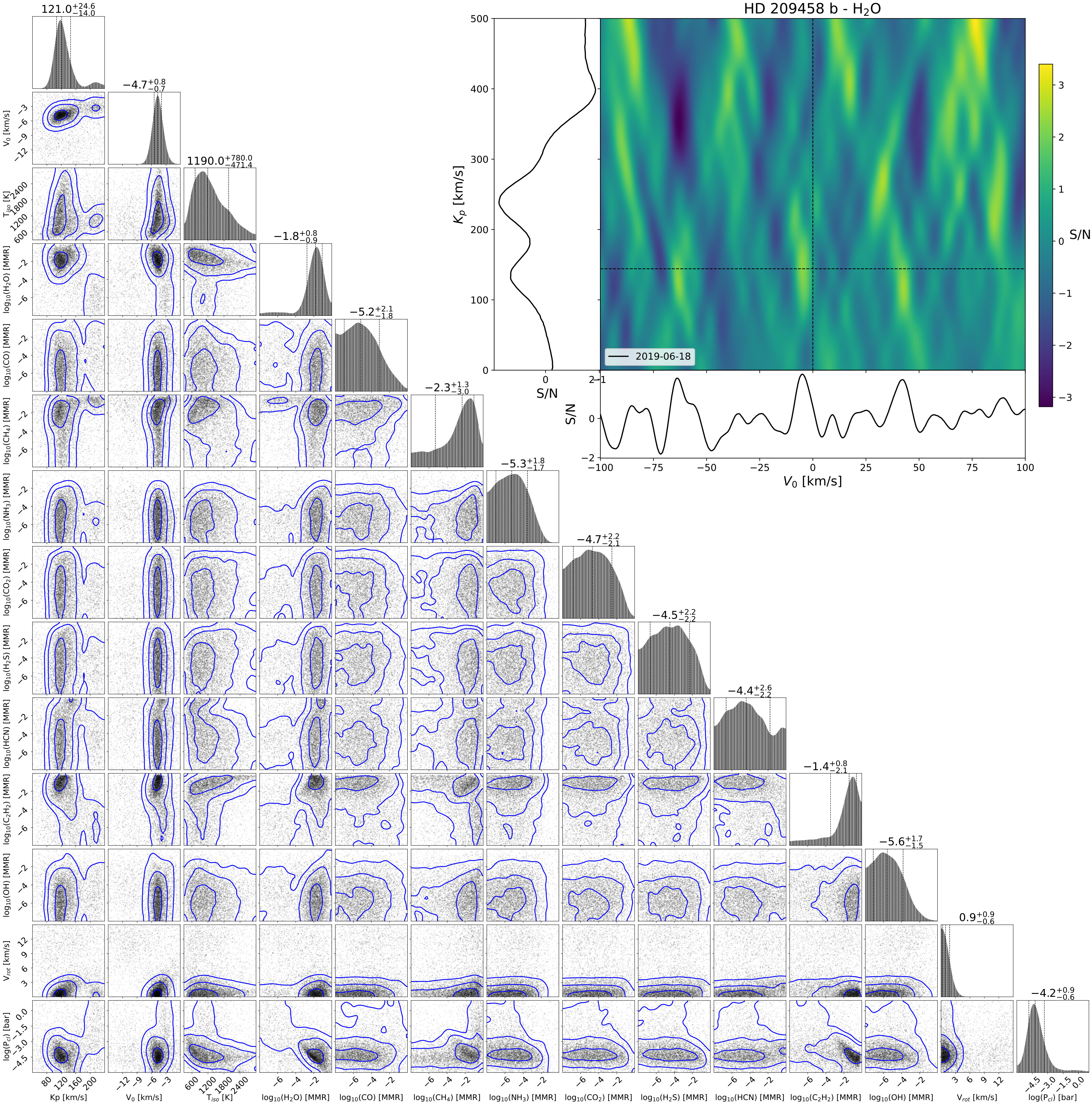}\hfill
  \includegraphics[width=0.395\linewidth,trim=0cm 0cm 0cm 0cm,clip]{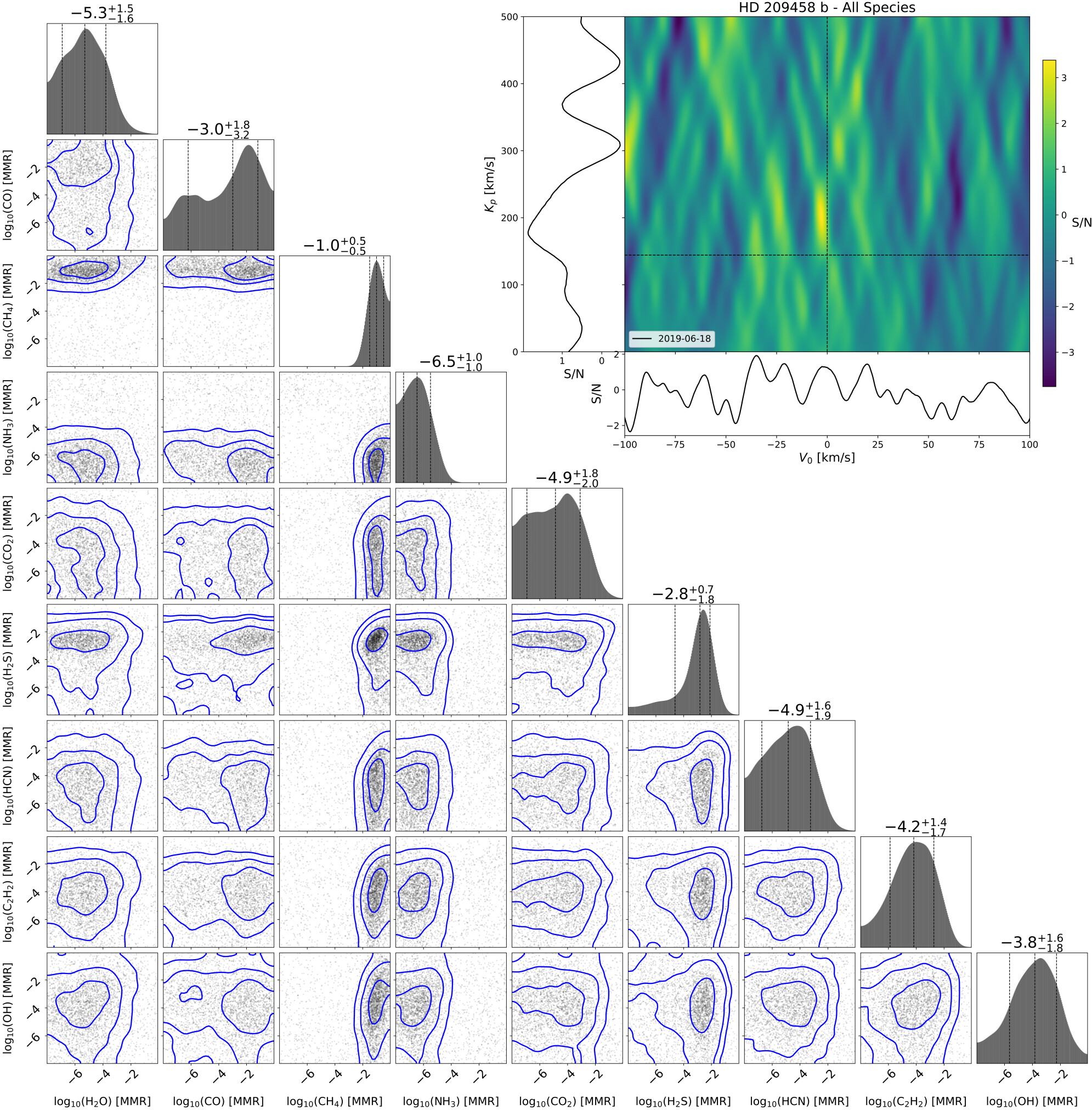}
  \caption{Corner plot from the \textit{Full} (left) and \textit{Upper Limits} (right) Runs with associated CCF maps for HD\,209458\,b. The blue contours represent the 1-, 2-, and 3-$\sigma$ confidence intervals. \textbf{Left:} the posteriors converged toward a signature compatible with the planetary signal (black cross in the CCF), yet corresponding to a 2.8-$\sigma$ and therefore inconclusive signal in the CCF map computed from the best \ce{H2O} abundance found. The projected CCF at the expected planetary Kp and V$_0=0$\,\kms (black cross) is shown in the left and lower CCF figure's panels for each transit date and for their combination (solid black). \textbf{Right:} posteriors and CCF map obtained from the \textit{Upper Limits Run}, i.e. by fixing the thermal profile to Exo-REM's predictions with a cloud-free model. The CCF map was obtained using the best abundances, including all species \red{and using the Exo-REM's thermal profile}, with the lower and left panels showing the CCF projected along the theoretical planet's position (black cross).}
  \label{fig:hd209}
\end{sidewaysfigure*} 

\begin{sidewaysfigure*}
  \centering
  \includegraphics[width=0.595\linewidth,trim=0cm 0cm 0cm 0cm,clip]{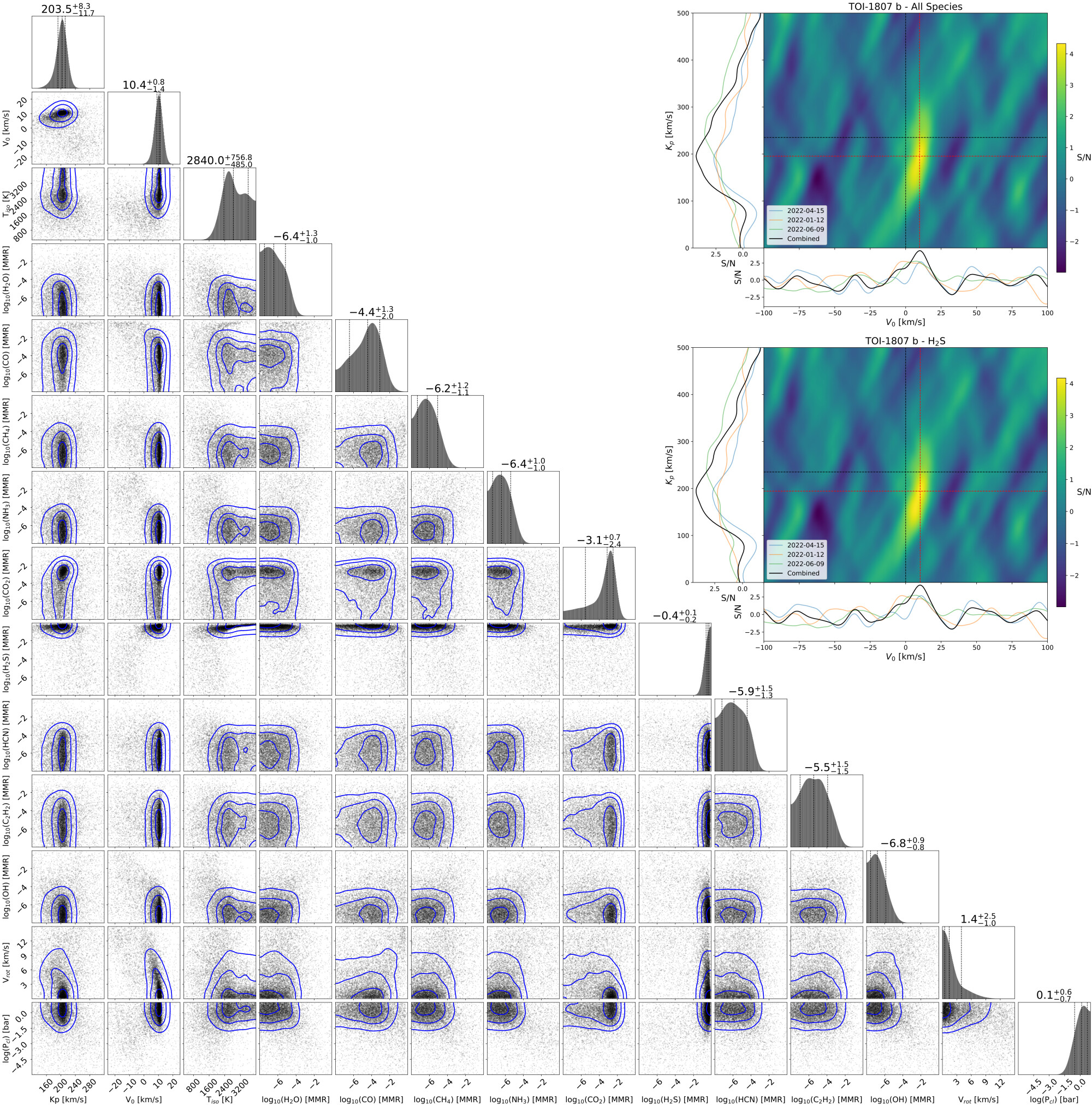}\hfill
  \includegraphics[width=0.395\linewidth,trim=0cm 0cm 0cm 0cm,clip]{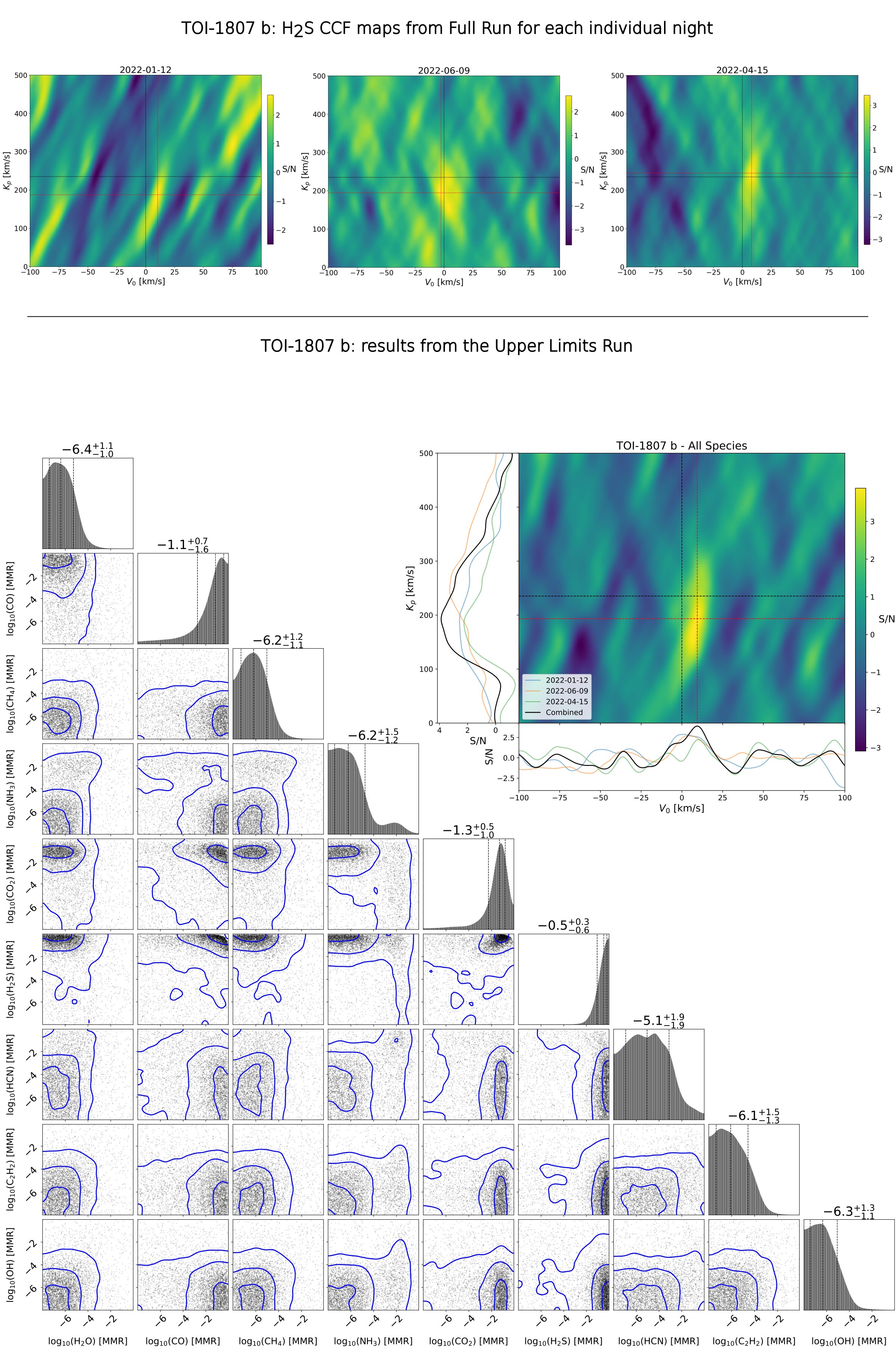}
  \caption{Corner plot from the \textit{Full} (left \red{and upper right}) and \textit{Upper Limits} (\red{lower} right) Runs with associated CCF maps for TOI-1807\,b. The blue contours represent the 1-, 2-, and 3-$\sigma$ confidence intervals. \textbf{Left:} the posteriors converged toward a signature at a lower Kp and higher V$_0$ than the expected planetary signal (black cross in the CCF), corresponding to a 4.4-$\sigma$ tentative signature when computing a CCF map including all species and with all parameters fixed to their best values (upper CCF figure). This signal is dominated by its \ce{H2S} contribution, with a \red{4.2}-$\sigma$ signature in the CCF map computed from an \ce{H2S} only model (lower CCF figure). The projected CCF at the best Kp and V$_0$\,\kms found by the retrieval (red cross) is shown in the left and lower CCF figures' panels for each transit date and for their combination (solid black). \red{\textbf{Upper Right:} CCF maps corresponding to the \ce{H2S}-\textit{Full Run} solution for each individual transit night.} \textbf{Lower Right:} posteriors and CCF map obtained from the \textit{Upper Limits Run}, i.e. by fixing the thermal profile to Exo-REM's predictions with a cloud-free model. The CCF map was obtained using the best abundances, including all species \red{and using the Exo-REM's thermal profile}, with the lower and left panels showing the CCF projected along the theoretical planet's position (black cross). The signal found in the \textit{Full Run} case is still visible at a lower \snr.}
  \label{fig:toi1807}
\end{sidewaysfigure*}

\end{appendix}

\end{document}